\documentclass[notitlepage,prd,showpacs,superscriptaddress,nofootinbib,floatfix,showkeys,10pt]{revtex4-1}
\usepackage{graphicx}
\usepackage{amsmath}
\usepackage{bm}
\usepackage{yhmath}
\usepackage{mathtools}
\usepackage{wasysym}
\usepackage{subfigure}
\usepackage{xcolor}
\definecolor{CherryRed}{rgb}{.65,0,.2}
\definecolor{RubyRed}{rgb}{.88,0.07,.3}
\definecolor{CralRed}{rgb}{1,0.25,.25}
\definecolor{CobaltBlue}{rgb}{0,0.28,.67}
\definecolor{RoyalBlue}{rgb}{0.25,0.41,.88}
\definecolor{EmeraldGreen}{rgb}{0.31,0.78,.47}
\usepackage{cases}
\usepackage[colorlinks=true, linkcolor=RubyRed,citecolor=CralRed,urlcolor=RubyRed,linkcolor=RubyRed]{hyperref}
\usepackage{subfigure}
\usepackage{times}
\usepackage{booktabs,bm}
\usepackage{slashed}
\usepackage{amsfonts,amssymb,stmaryrd,latexsym,amsmath}
\usepackage{textcomp}
\usepackage{multirow}
\usepackage{cancel}
\usepackage{array}
\usepackage{hyperref}
\usepackage{pgf}
\usepackage{orcidlink}
\usepackage{cleveref}

\def\SU3{{\text{SU(3)}_{\rm F}}}

\def \pcs4338{{P_{\psi s}^\Lambda(4338)^0}}

\begin{document}
	
	\title{\textcolor{CobaltBlue}{XXXXXXXXX}}

	\date{\today}

	
\title{Study of the semileptonic decay $\Lambda \to p\,\ell\,\bar{\nu}_{\ell}$ in QCD 
}
\author{M. Ahmadi\orcidlink{0009-0006-4046-2121}}
\email{masoumehahmadi@ut.ac.ir}
\affiliation{Department of Physics, University of Tehran, North Karegar Avenue, Tehran 14395-547, Iran}
\author{Z.~Rajabi Najjar$^{a}$\orcidlink{0009-0002-2690-334X}}
	\email{rajabinajar8361@ut.ac.ir }
\affiliation{Department of Physics, University of Tehran, North Karegar Avenue, Tehran 14395-547, Iran}
\author{K. Azizi\orcidlink{0000-0003-3741-2167}}
\email{kazem.azizi@ut.ac.ir}
\thanks{Corresponding author}
\affiliation{Department of Physics, University of Tehran, North Karegar Avenue, Tehran
14395-547, Iran}
\affiliation{Department of Physics, Dogus University, Dudullu-\"{U}mraniye, 34775
Istanbul,  T\"{u}rkiye}

\date{\today}
	
\preprint{}
	
\begin{abstract}
We conduct a comprehensive study of the semileptonic decay process 
\(\Lambda \to p\,\ell\,\bar{\nu}_{\ell}\), focusing on the determination of  all six 
vector and axial-vector form factors that govern the low-energy hadronic matrix 
elements of the underlying theory. These invariant form factors constitute the 
essential inputs for describing the decay, and their dependence on the momentum 
transfer \(q^{2}\) is analyzed across the entire physical kinematic region. 
To parametrize the \(q^{2}\)dependence, we adopt both the \(z\)-expansion 
formalism and a polynomial fitting approach. Utilizing parametrizations, 
we compute the exclusive decay widths for both the electron and 
muon channels, and subsequently extract the corresponding 
branching ratios. Furthermore, we evaluate the ratio of decay widths between 
the muon and electron channels defined as 
$R^{\mu e} \equiv \frac{\Gamma(\Lambda \to p\,\mu\,\bar{\nu}_{\mu})}{\Gamma(\Lambda \to p\,e\,\bar{\nu}_{e})}$  obtaining 
\(R^{\mu e} = 0.196^{+0.009}_{-0.012}\) from the polynomial fit and 
\(R^{\mu e} = 0.174^{+0.002}_{-0.005}\) from the \(z\) expansion. While both 
ratios are compatible with previously reported values in the literature, 
the result from the \(z\) expansion exhibits particularly strong agreement with 
the averages reported by the Particle Data Group.

\end{abstract}
	
	
	
\maketitle
	
\renewcommand{\thefootnote}{\#\arabic{footnote}}
\setcounter{footnote}{0}

	\maketitle
	\renewcommand{\thefootnote}{\#\arabic{footnote}}
	\setcounter{footnote}{0}


\section {Introduction}\label{sec:one}

Hyperon semileptonic decays (HSDs) provide a powerful framework for investigating both strong and weak interactions in the light-baryon sector. The strong interaction is manifested through parton hadronization, while the weak interaction drives flavor-changing processes. Furthermore, hyperon decays, as baryonic systems containing strange quarks, offer promising opportunities for probing sources $CP$ violation. Beyond their role in testing the Standard Model (SM), HSDs allow detailed studies of the interplay between weak and strong dynamics through analyses of form factors, branching ratios, angular correlations, and decay rates. Although their branching ratios are typically of order $\mathcal{O}(10^{-4})$, hyperon semileptonic decays remain essential channels for investigating low-energy strong interaction dynamics and for providing sensitive probes of potential new physics~\cite{Donoghue:1985ww,Gaillard:1984ny}. The Cabibbo-Kobayashi-Maskawa (CKM) matrix element $|V_{us}|$ determines the strength of weak transitions between strange and up quarks. In view of this discrepancy, independent determinations of $|V_{us}|$ are essential. Its current value, however, shows  tension between different determination methods, including extractions from kaon and $\tau$ decays, as well as from the CKM first-row unitarity relation. Since the investigation of hyperon beta decay is essential for understanding the internal structure of hadrons, the decay process $\Xi^0 \to \Sigma^+ e^- \bar{\nu}_e$ is of particular interest. It is analogous to the well-studied neutron decay $n \to p \, e^- \, \bar{\nu}_e$, with the distinction that, in the hyperon case, the valence $d$ quarks are replaced by $s$ quarks in both the initial and final baryon states~\cite{KTeVE832E799:1999tte,KTeV:2001djr}.  
Charged-current semileptonic decays of $b$ hadrons are of considerable phenomenological importance, as they offer a clean framework for constraining the CKM matrix elements $V_{ub}$ and $V_{cb}$. Reference ~\cite{Rui:2025bsu} reports the first perturbative QCD determination of the $\Xi_b \to (\Lambda, \Sigma)$ transition form factors at leading order in $\alpha_s$, which govern the Cabibbo-suppressed semileptonic decays $\Xi_b \to (\Lambda, \Sigma)\,\ell \nu_{\ell}$ with $\ell = e, \mu, \tau$. 
 In Ref.~\cite{Gonzalez-Alonso:2016etj}, within the framework of the Standard Model effective field theory at the electroweak scale, the authors introduced a global, model-independent analysis of potential new-physics contributions in the $D \to u\,\ell\,\nu$ transitions ($D = d, s$; $\ell = e, \mu$).
 
  In this context, semileptonic hyperon decays, such as $\Lambda \to p\,\ell\,\bar{\nu}_{\ell}$, provide an alternative and complementary approach for the extraction of $|V_{us}|$. In fact, precision measurements of $|V_{us}|$ from HSDs not only constitute stringent tests of the SM but also have the potential to reveal signs of physics beyond it~\cite{Cabibbo:2003cu,Cabibbo:2003ea}. The ratio of decay rates between the semimuonic and semielectronic channels, $R^{\mu e} \equiv \Gamma(B_1 \to B_2 \mu^- \bar{\nu}_\mu)/\Gamma(B_1 \to B_2 e^- \bar{\nu}_e)$, has been studied extensively in the literature. For the decay $\Lambda \to p$, Ref.~\cite{Chang:2014iba} predicts $R^{\mu e} = 0.153^{+0.008}_{-0.008}$, whereas Ref.~\cite{BESIII:2021ynj} finds $R^{\mu e} = 0.178^{+0.028}_{-0.028}$ by computing $\mathcal{B}(\Lambda \to p \mu^- \bar{\nu}_\mu) = (1.48^{+0.21}_{-0.21} \pm 0.08) \times 10^{-4}$ and combining it with the precisely measured branching ratio of the electron-channel decay, $\mathcal{B}(\Lambda \to p e^- \bar{\nu}_e)$.
  
Moreover, in HSDs the effects of SU(3) symmetry breaking have been investigated using chiral perturbation theory, with studies examining corrections with $f_{1}(0)$ beyond the first-order approximation~\cite{Villadoro:2006nj,Lacour:2007wm,Faessler:2008ix}. These investigations are further complemented by detailed analyses of other form factors relevant to HSDs within the framework of the $1/N_c$ expansion~\cite{Flores-Mendieta:2004cyh}.

Lattice QCD offers a powerful nonperturbative framework for investigating SU(3) symmetry-breaking effects in hyperon semileptonic decays~\cite{Guadagnoli:2006gj,Cooke:2012xv,Sasaki:2008ha,Sasaki:2012ne,Sasaki:2017jue}, and it has facilitated precise determinations of the form factors governing semileptonic baryon decays~\cite{Detmold:2015aaa,Meinel:2017ggx,Meinel:2021rbm,Farrell:2025gis,Bacchio:2025auj}. This includes pioneering lattice studies of the decays $\Xi_c \to \Xi \,\ell^{+} \nu_{\ell}$~\cite{Farrell:2025gis} and $\Lambda \to p \,\ell \,\bar{\nu}_{\ell}$, for which the ratio $R^{\mu e}$ has been determined as $0.1880(83)$~\cite{Bacchio:2025auj}.

 QCD sum rules offer a complementary, model-independent approach grounded in the first principles of QCD. This method employs three-point correlation functions constructed from interpolating currents, with nonperturbative contributions encoded in vacuum condensates. By matching the QCD and phenomenological representations of these correlators via quark-hadron duality, weak transition form factors can be systematically extracted. Originally developed by Ioffe and Smilga~\cite{Ioffe:1982ia} and Nesterenko and Radyushkin ~\cite{Nesterenko:1982gc} for the study of pion electromagnetic form factors , this approach represents the decay amplitude in two complementary ways: in terms of hadronic parameters, including baryon masses, residues, and transition form factors, and in terms of the underlying QCD degrees of freedom, such as quark masses, the strong coupling constant, and vacuum condensates.
 The transition form factors are then extracted using quark-hadron duality, Borel transformations, and continuum subtraction, and subsequently employed to compute semileptonic decay widths for different lepton channels.

This framework has been widely applied to baryon semileptonic decays~\cite{Dai:1996xv,Huang:1998rq,MarquesdeCarvalho:1999bqs,Huang:1998ek,Gan:2010hw,Dag:2010jr,Wang:2012hu,Zhao:2021sje,Xing:2021enr,Zhang:2023nxl,Najjar:2024ngm,Neishabouri:2024gbc,Amiri:2025gcf,Zhang:2024ick,Aliev:2010uy,Azizi:2009wn,Azizi:2018axf}. In particular, Ref.~\cite{Zhang:2024ick} analyzes the semileptonic decays $\Lambda \to p\,\ell\,\bar{\nu}_{\ell}$, $\Sigma^{0} \to p\,\ell\,\bar{\nu}_{\ell}$, $\Xi^{-} \to \Lambda\,\ell\,\bar{\nu}_{\ell}$, and $\Xi^{-} \to \Sigma^{0}\,\ell\,\bar{\nu}_{\ell}$ within the QCD sum-rule framework. Among these, the decay $\Lambda \to p\,\ell\,\bar{\nu}_{\ell}$ is of particular interest, providing access to CKM matrix elements, tests of SU(3) symmetry breaking, and baryonic form factors that probe nonperturbative QCD dynamics. Its study of different lepton channels also allows tests of lepton flavor universality and potential signals of physics beyond the SM.

 The paper is organized as follows: In Sec.~II, we describe the QCD sum-rule framework and derive the form-factor sum rules from the corresponding correlation function. Section~III presents the numerical analysis, and in Sec.~IV we report the results for the semileptonic decay widths in all lepton channels. Our conclusions are given in Sec.~V, and further technical details are provided in the Appendix.

\section { THEORETICAL FRAMEWORK }\label{sec:two}
\begin{figure}[h!] 
		\includegraphics[totalheight=6cm,width=9cm]{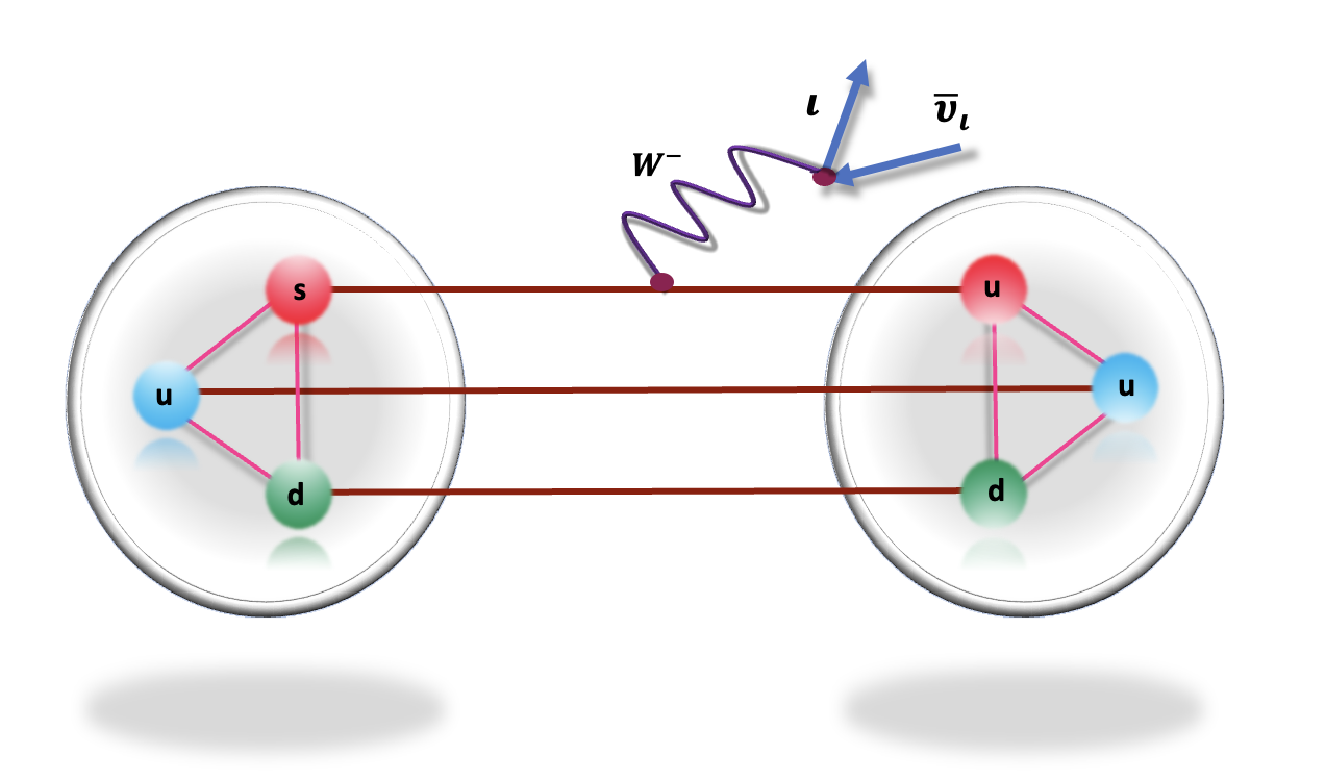}
		\caption{Illustration of the semileptonic decay  $ \Lambda \rightarrow p ~{\ell}\bar\nu_{\ell}$.}\label{Fig:current}
	\end{figure}

In the semileptonic  decay process $ \Lambda \rightarrow p ~{\ell}\bar\nu_{\ell}$ , the two quarks $u$ and $d$ present in the initial baryon remain unchanged and act merely as spectators,  while the main transition occurs by the quark $s$ through the weak interaction. As illustrated in Fig.~\ref{Fig:current}, during this transition, the $s$ quark decays through the vector boson  $W^-$, producing  a quark $u$,  $\ell$, and $\bar\nu_{\ell}$. This process is described by the following current:	
\begin{eqnarray}
{\cal
		J}_{\mu}^{tr}=\bar u \gamma_\mu(1-\gamma_5) s.
		\end{eqnarray}
In the above relation, the factor $\gamma_\mu(1-\gamma_5)$ represents a combination of vector and axial-vector currents within the framework of the weak interaction. The effective Hamiltonian corresponding to this transition, which accounts for the weak interaction between quarks and leptons, is given by:
\begin{eqnarray}\label{Heff}
	{\cal H}_{eff} =
	\frac{G_F}{\sqrt2} V_{us} ~\bar u \gamma_\mu(1-\gamma_5) s \bar{\ell}\gamma^\mu(1-\gamma_5) \nu_{\ell}.
\end{eqnarray}

Now, to obtain the transition amplitude for the decay process, we employ the method of sandwiching the effective Hamiltonian between the initial state $\Lambda$ and the final state $p$. The transition amplitude for this decay is then given as follows:
\begin{eqnarray}\label{amp}
	M=\langle p \vert{\cal H}_{eff}\vert \Lambda \rangle 
=\frac{G_F}{\sqrt2} V_{us}\bar{\ell}~\gamma^\mu(1-\gamma_5) \nu_{\ell} \langle  p \vert \bar u \gamma_\mu(1-\gamma_5)s\vert \Lambda \rangle,
\end{eqnarray}
where $G_F$ is the Fermi constant,~ and $V_{us}$ is the corresponding element of the CKM matrix. As can be seen, the transition amplitude contains both the vector components $\bar u \gamma_\mu s$ and the axial-vector $\bar u \gamma_\mu\gamma_5s$ each of which contains information about three form factors based on their  Lorentz structure. Taking into consideration  Lorentz invariance and parity, the transition matrices are defined in terms of six form factors as follows: 

\begin{eqnarray}\label{Cur.with FormFac.}
	&&\langle p(p',s')|V^{\mu}|\Lambda (p,s)\rangle = \bar
	u_p(p',s') \Big[F_1(q^2)\gamma^{\mu}+F_2(q^2)\frac{p^{\mu}}{m_\Lambda}
	+F_3(q^2)\frac{p'^{\mu}}{m_p}\Big] u_\Lambda(p,s), \notag \\
	&&\langle p(p',s')|A^{\mu}|\Lambda (p,s)\rangle = \bar u_p(p',s') \Big[G_1(q^2)\gamma^{\mu}+G_2(q^2)\frac{p^{\mu}}{m_\Lambda}+G_3(q^2)\frac{p'^{\mu}}{m_p}\Big]
	\gamma_5 u_\Lambda(p,s).
\end{eqnarray}
Here, the quantities $ F_i $ and $G_i$ ($ i$=1,2,3) are introduced as form factors corresponding to the vector and axial parts of the transition current, respectively. The initial baryon carries four-momentum $p$, and the final baryon is described by the $p'$. Their difference $q = p - p'$ represents the momentum transferred to the lepton pair.  The Dirac spinor states of these baryons are denoted by $u_{\Lambda}$ and $u_P$. Form factors are among the most important parameters, which can be calculated using the QCD sum-rule approach. The procedure typically begins with the choice of a suitable correlation function, and to determine these functions, the general form of the three-point correlation function is written down as follows:
\begin{eqnarray}\label{CorFunc}
	\Pi_{\mu}(p,p^{\prime},q)&=&i^2\int d^{4}x e^{-ip\cdot x}\int d^{4}y e^{ip'\cdot y}  \langle 0|{\cal T}\{{\cal J}^{p}(y){\cal
		J}_{\mu}^{tr,V(A)}(0) \bar {\cal J}^{\Lambda}(x)\}|0\rangle.
\end{eqnarray}
In the above formula,  $\mathcal{T}$ represents the time ordering of operators. As previously said,  ${\cal
	J}_{\mu}^{tr,V(A)}(0)$ refers to the transition current associated with the semileptonic decay. Furthermore,  $ {\cal J}^{\Lambda}(x)$ and ${\cal J}^{p}(y) $ are the interpolating currents related to the initial and final particles. In this process, the initial state is the $\Lambda$, which is a hyperon with spin 1/2, and the final state is the proton, $p$. The general form of the interpolating currents for these two baryons can be written as follows:
\begin{eqnarray}\label{cur1}
{\cal J}^{\Lambda}(x) &=& 
2 \sqrt{\tfrac{1}{6}} \, \varepsilon^{abc} \Big\{ 
 2 \big(u^{aT} C d^b\big) \gamma_5 s^c 
 + \big(u^{aT} C s^b\big) \gamma_5 d^c 
 - \big(d^{aT} C s^b\big) \gamma_5 u^c \nonumber \\
&& \hspace{1.1cm}
+\, 2\beta \big(u^{aT} C \gamma_5 d^b\big) s^c 
+ \beta \big(u^{aT} C \gamma_5 s^b\big) d^c 
 - \beta \big(d^{aT} C \gamma_5 s^b\big) u^c 
\Big\} ,
\end{eqnarray}
and
\begin{align}\label{eq:CorrF2}
{\cal J}^{p}(x) = 2\, \epsilon^{abc} \Big\{
   \big(u^{a^T} C d^b\big)\gamma_5 u^c
   + \beta \big(u^{a^T} C \gamma_5 d^b\big) u^c
\Big\},
\end{align}
where quark fields are functions of the space-time $x  $.

The color indices are denoted by $a$, $b$, and $c$, the charge conjugation operator by $C$, and $\beta$ represents an auxiliary parameter. Within the QCD sum-rule framework, the correlation function is evaluated using two independent representations: the hadronic description and the QCD representation. By comparing and matching these two representations, the transition form factors can be systematically extracted. This procedure involves the use of dispersion integrals together with the assumption of quark-hadron duality.  In order to  suppress contributions from higher excited states, the Borel transformation is applied in combination with continuum subtraction.

\subsection{Hadronic side} 

To evaluate the correlation function on the hadronic side, the calculations are performed in the timelike region. At this stage, it is essential to introduce complete sets of intermediate states that share the same quantum numbers as the initial baryon $\Lambda$ and the final baryon $p$. By inserting these states and integrating over the four-dimensional coordinates $x$ and $y$, the hadronic representation of the correlation function is obtained as
follows:

\begin{eqnarray} \label{PhysSide}
		\Pi_{\mu}^{Had.}(p,p',q)=\frac{\langle 0 \mid {\cal J}^{p} (0)\mid p(p') \rangle \langle p (p')\mid
			{\cal J}_{\mu}^{tr,V(A)}(0)\mid \Lambda(p) \rangle \langle \Lambda(p)
			\mid\bar {\cal J}^{\Lambda}(0)\mid
			0\rangle}{(p'^2-m_{p}^2)(p^2-m_{\Lambda}^2)}+\cdots.
	\end{eqnarray}

Here, the first term corresponds to the ground-state contribution, while the remaining terms indicated by $\cdots$  represent the excited resonances and the continuum part. To specify the matrix elements appearing in this relation, the following definitions are applied:
\begin{eqnarray}\label{MatrixElements}
	&&\langle 0|{\cal J}^{p}(0)|p(p')\rangle =
	\lambda_{p} u_{p}(p',s'), \notag \\
	&&\langle \Lambda(p)|\bar {\cal J}^{\Lambda}(0)| 0 \rangle =
	\lambda^{\dag}_{\Lambda}\bar u_{\Lambda}(p,s),
\end{eqnarray}
where, $\lambda_{\Lambda} $ and $\lambda_{p}$ denote the residues of the initial and final states, respectively. For the $\Lambda$ (as the initial state), by fitting the corresponding residue plot given in Ref. \cite{Aliev:2002ra}, it  is obtained as follows and used in the calculations:
\begin{eqnarray}
\lambda_{\Lambda}^2(\beta)&=&e^{\frac{m_{\Lambda}^2}{M^{2}}}[9 - 23.73\beta+340.6\beta^{2}]\times10^{-6} ~~(\mathrm{GeV^6}).
\end{eqnarray}

Similarly, for the proton (as the final state), the residue is defined as follows \cite{Azizi:2015fqa}:

\begin{eqnarray}
\lambda_{p}^2(\beta)&=&e^{\frac{m_{N}^2}{M^{\prime^2}}}
[2.28-3.20\beta+64.10\beta^{2}]\times10^{-6} ~~(\mathrm{GeV^6}),
\end{eqnarray}
where  the quantities $ M^2$ and $ M'^2$ are considered  the Borel parameters. These parameters are determined in the numerical analysis by defining appropriate windows.  Moreover, the Dirac spinors corresponding to these two baryons, namely,  $u_{\Lambda}(p,s)$ and $u_{P}(p',s')$, satisfy the following mathematical relations:
 \begin{eqnarray}\label{Spinors}
	\sum_{s'} u_{p} (p',s')~\bar{u}_{p}
	(p',s')&=&\slashed{p}~'+m_p,\notag \\
	\sum_{s} u_{\Lambda}(p,s)~\bar{u}_{\Lambda}(p,s)&=&\slashed
	p+m_{\Lambda}.
\end{eqnarray}
 By substituting these relations into Eq. (\ref{PhysSide}), the general form of the correlation function on the hadronic side is obtained: 
\begin{eqnarray} \label{PhysSidetotal}
	\Pi_{\mu}^{Had.}(p,p',q)&=&\frac{\lambda_{p}\lambda^{\dag}_{\Lambda}(\slashed{p}~'+m_p)(\slashed{p}~+m_{\Lambda})}{(p'^2-m_p^2)(p^2-m_{\Lambda}^2)}\Bigg( \Big[F_1(q^2)\gamma^{\mu}+F_2(q^2)\frac{p^{\mu}}{m_{\Lambda}}
		+F_3(q^2)\frac{p'^{\mu}}{m_p}\Big]\notag\\ &-&\Big[G_1(q^2)\gamma^{\mu}\gamma_5+G_2(q^2)\frac{p^{\mu}\gamma_5}{m_{\Lambda}}+G_3(q^2)\frac{p'^{\mu}\gamma_5}{m_p}\Big]
		\Bigg)+\cdots.
\end{eqnarray}
 To eliminate the contributions of the excited states and the continuum region, the double Borel transformation method is employed: 
 \begin{eqnarray}\label{BorelQCD2}
	\mathbf{\widehat{B}}\frac{1}{(p^{2}-s)^m} \frac{1}{(p'^{2}-s')^n}\longrightarrow (-1)^{m+n}\frac{1}{\Gamma[m]\Gamma[n]} \frac{1}{(M^2)^{m-1}}\frac{1}{(M'^2)^{n-1}}e^{-s/M^2} e^{-s'/M'^2}.
\end{eqnarray} 
Within the Borel scheme, the following relation  holds:
 \begin{eqnarray}\label{Physical Side structures}
&&\mathbf{\widehat{B}}~\Pi_{\mu}^{\mathrm{Phys.}}(p,p',q)=\lambda_{ \Lambda}\lambda^{\dag}_{P}~e^{-\frac{m_{ \Lambda}^2}{M^2}}
~e^{-\frac{m_p^2}{M'^{2}}}\Bigg[F_{1}\bigg(m_{\Lambda} m_p \gamma_{\mu}+m_{ \Lambda} \slashed{p}' \gamma_{\mu}+m_p\gamma_{\mu}\slashed {p}+\slashed {p}'\gamma_\mu\slashed {p}\bigg)+\notag\\
&&F_2\bigg(\frac{m_p}{m_{\Lambda}}p_\mu\slashed {p}+\frac{1}{m_{ \Lambda}}p_{\mu}\slashed {p}' \slashed {p}+m_pp_\mu +p_\mu\slashed {p}'\bigg)+ F_3\bigg(\frac{1}{m_p} p'_{\mu} \slashed {p}' \slashed{p}+p'_\mu\slashed {p}'+p'_\mu\slashed {p}+m_{ \Lambda}p'_\mu\bigg)-\notag\\
&& G_1\bigg(m_{\Lambda} m_p \gamma_{\mu}\gamma_{5}+m_{ \Lambda}\slashed {p}'\gamma_\mu\gamma_5-m_p\gamma_\mu\slashed {p}\gamma_5-\slashed {p}'\gamma_\mu\slashed {p}\gamma_5\bigg)- G_2\bigg(p_\mu\slashed {p}'\gamma_5+m_pp_\mu\gamma_5-\frac{m_p}{m_{ \Lambda}}p_\mu\slashed {p}\gamma_5-\frac{1}{m_{ \Lambda}} p_{\mu} \slashed {p}' \slashed
{p}\gamma_{5}\bigg)\notag\\
&&-G_3\bigg(\frac{m_{ \Lambda}}{m_p}p'_\mu\slashed {p}'\gamma_5+m_{ \Lambda}p'_\mu\gamma_5-\frac{1}{m_p} p'_{\mu}
\slashed {p}'\slashed{p}\gamma_{5}-p'_\mu\slashed {p}\gamma_5\bigg)\Bigg]+\cdots~.
\end{eqnarray}
 
	 	 \subsection{QCD side} 

According to the QCD sum-rule approach, calculating the correlation function on the QCD side requires its probe in the deep Euclidean region using the operator product expansion (OPE). For this purpose, first the interpolating currents of the initial and final states [Eqs. (\ref{cur1}) and (\ref{eq:CorrF2})] are substituted into Eq. (\ref{CorFunc}). Then, by applying Wick’s theorem, all possible contractions among the quark fields are computed, leading to an explicit form of the correlation function that depends on the propagators of the light quarks:
\begin{eqnarray} \label{ term}
	&&\Pi^{OPE}_{\mu}=i^2 \int d^4x e^{-ipx}\int d^4y e^{ip'y} \frac{2}{\sqrt{3}} \epsilon_{a'b'c'} \epsilon_{abc}\Bigg\{-2 \gamma_5~ S^{c a'}_u(y-x) S'^{b b'}_d(y-x) S^{a i}_u(y) \gamma_\mu (1-\gamma_5) S^{i c'}_s(-x) \gamma_5\notag\\
	&&- Tr[S'^{b b'}_d(y-x) S^{a i}_u(y) \gamma_\mu(1-\gamma_5) S^{ i a'}_s(-x)] ~\gamma_5 S^{c c'}_u(y-x)\gamma_5-
	\beta Tr[S'^{b b'}_d(y-x)  \gamma_5 S^{a i}_u(y)\gamma_\mu (1-\gamma_5) S^{i a'}_s(-x)] \notag\\
	&&\gamma_5 S^{c c'}_u(y-x)- \beta  S'^{b c'}_d(y-x)\gamma_5~ S^{i b'}_s(-x) \gamma_5 S^{a a'}_u(y-x) \gamma_\mu (1-\gamma_5) S^{c i}_u(y) - Tr[S'^{a a'}_u(y-x) S^{b b'}_d(y-x)] \gamma_5 S^{c i}_u(y) \notag\\
	&&\gamma_\mu(1-\gamma_5) S^{i c'}_s(-x) ~\gamma_5-\gamma_5~ S^{c i}_u(y) \gamma_\mu (1-\gamma_5) S^{i b'}_s(-x) S'^{a a'}_u(y-x)  S^{b c'}_d(y-x)\gamma_5  +2\beta \gamma_5 S^{ic'}_s(-x) \gamma_\mu (1-\gamma_5) S^{c i}_u(y)  \notag\\
	&&Tr[  S'^{b b'}_d(y-x) \gamma_5 S^{a a'}_u(y-x) ]-\beta   S'^{b b'}_d(y-x) \gamma_5 S^{i a'}_s(-x)\gamma_5 S^{a c'}_u(y-x)  \gamma_\mu(1-\gamma_5) ~S^{c i}_u(y) - \gamma_5 S^{c a'}_u(y-x) S'^{i b'}_s(-x)  \notag\\
	&&  (1-\gamma_5)\gamma_\mu S'^{a i}_u(y) S^{b c'}_d(y)  \gamma_5  + \gamma_5 S^{c i}_u(y)    \gamma_\mu(1-\gamma_5) S^{i a'}_s(-x) S'^{b b'}_d(y-x)  S^{a c'}_u(y-x)  \gamma_5 -\beta S'^{i b'}_s(-x)  \gamma_5 S^{c a'}_u(y-x)  \gamma_5 \notag\\
	&&S^{b c'}_d(y-x)  \gamma_\mu(1-\gamma_5)  ~S^{a i}_u(y) +\beta   S'^{a a'}_u(y-x) \gamma_5 S^{b c'}_d(y-x)  \gamma_5 S^{i b'}_s(-x)  \gamma_\mu (1-\gamma_5)   ~S^{c i}_u(y) + \beta Tr[S'^{b b'}_d(y-x)  \gamma_5  S^{a i}_u(y)  \gamma_\mu\notag\\
	&& (1-\gamma_5) S^{i a'}_s(-x) ] \gamma_5 S^{c c'}_u(y-x)    - 2 \beta Tr[S'^{b b'}_d(y-x)  \gamma_5  S^{a i}_u(y-x) ]  \gamma_5  S^{i c'}_u(y)  \gamma_\mu (1-\gamma_5)   ~S^{c a'}_s(-x)   - \beta^2 \gamma_5 S'^{a i}_u(y)    \notag\\
	&&\gamma_5 S^{i b'}_s(-x)  \gamma_\mu (1-\gamma_5) S^{b c'}_d(y-x)    S^{c a'}_u(y-x)  - 2 \beta^2 Tr[\gamma_5 S'^{b b'}_d(y-x) \gamma_5   S^{a a'}_u(y-x) ]\gamma_\mu  (1-\gamma_5)    S^{i c'}_s(-x) S^{c i}_u(y)  \notag\\
	&&+2 \beta  S'^{b b'}_d(y-x)  \gamma_5  S^{a i}_u(y-x)  \gamma_5  S^{i c'}_s(-x) \gamma_\mu   (1-\gamma_5)  S^{c a'}_u(y)     +  \beta S'^{b b'}_d(y-x)  \gamma_5 S^{i a'}_s(-x) \gamma_5  S'^{c i}_u(y)   \gamma_\mu (1-\gamma_5)   S^{a c'}_u(y-x)    \notag\\
	&&-  \beta^2  Tr[ \gamma_5  S'^{b b'}_d(y-x)  \gamma_5  S^{i a'}_s(-x)  \gamma_\mu   (1-\gamma_5) S^{a i}_u(y)]  S^{c c'}_u(y-x)  + \beta^2  \gamma_5 S'^{a a'}_u(y-x)  \gamma_5 S^{i b'}_s(-x)  \gamma_\mu (1-\gamma_5) S^{b c'}_d(y-x)       \notag\\
	&& S^{c i}_u(y)  +  \beta  S'^{b c'}_d(y-x) \gamma_5   S^{i b'}_s(-x) \gamma_5  S^{a i}_u(y) \gamma_\mu   (1-\gamma_5)  S^{c a'}_u(y-x)   -2 \beta  S'^{b b'}_d(y-x)\gamma_5 S^{a i}_u(y) \gamma_5   S^{i c'}_s(-x)       \notag\\
	&& \gamma_\mu(1-\gamma_5)  S^{c a'}_u(y-x)   + 2 \beta^2 \gamma_5  S'^{b b'}_d(y-x)  \gamma_5 S^{a i}_u(y)    \gamma_\mu (1-\gamma_5)   S^{ i c'}_s(-x)    S^{c a'}_u(y-x)  + \beta^2 \gamma_5 S'^{b b'}_d(y-x) \gamma_5  S^{i a'}_s(-x)     \notag\\
	&& \gamma_\mu (1-\gamma_5)  S^{a c'}_u(y-x) S^{c i}_u(y)  \Bigg\}.
\end{eqnarray}

In this relation,  $S_i$ is introduced as the light-quark propagator, and $S'=CS^TC$. To carry out the calculations in coordinate space, one needs to employ the expression of the light-quark propagators:
\begin{eqnarray}\label{LightProp}
	S_{q}^{ab}(x)&=&i\delta _{ab}\frac{\slashed x}{2\pi ^{2}x^{4}}-\delta _{ab}%
	\frac{m_{q}}{4\pi ^{2}x^{2}}-\delta _{ab}\frac{\langle\overline{q}q\rangle}{12} +i\delta _{ab}\frac{\slashed xm_{q}\langle \overline{q}q\rangle }{48}%
	-\delta _{ab}\frac{x^{2}}{192}\langle \overline{q}g_{}\sigma
	Gq\rangle+
	i\delta _{ab}\frac{x^{2}\slashed xm_{q}}{1152}\langle \overline{q}g_{}\sigma Gq\rangle \notag\\
	&-&i\frac{g_{}G_{ab}^{\alpha \beta }}{32\pi ^{2}x^{2}}\left[ \slashed x{\sigma _{\alpha \beta }+\sigma _{\alpha \beta }}\slashed x\right]-i\delta _{ab}\frac{x^{2}\slashed xg_{}^{2}\langle
		\overline{q}q\rangle ^{2}}{7776} -\delta _{ab}\frac{x^{4}\langle \overline{q}q\rangle \langle
		g_{}^{2}G^{2}\rangle }{27648}+\ldots,
\end{eqnarray}
where the quantities \(\langle \overline{q}q\rangle\), \(\langle G^2\rangle\), and \(\langle \overline{q}g_{}\sigma Gq\rangle\) represent quark-quark, gluon-gluon, and quark-gluon condensates, respectively.  After substituting these propagators in coordinate space into the correlation function, the QCD calculations proceed using processes  such as Fourier transformation, Feynman parametrization,  application of some mathematical identities, and performing  the resultant  integrals. The outcome of this procedure is an expression for the correlation function on the QCD side, which contains 24 independent Lorentz structures:
\begin{eqnarray}\label{Structures}
&&\Pi_{\mu}^{\mathrm{QCD}}(p,p',q)=\Pi^{\mathrm{QCD}}_{\slashed{p}' \gamma_{\mu}\slashed{p}}(p^{2},p'^{2},q^{2})~\slashed{p}' \gamma_{\mu}\slashed{p}+
\Pi^{\mathrm{QCD}}_{p_{\mu} \slashed {p}'\slashed {p}}(p^{2},p'^{2},q^{2})~p_{\mu} \slashed {p}'\slashed {p}+
\Pi^{\mathrm{QCD}}_{p_{\mu}' \slashed {p}'\slashed {p}}(p^{2},p'^{2},q^{2})~p_{\mu}' \slashed {p}'\slashed {p}+\Pi^{\mathrm{QCD}}_{p'_\mu\slashed {p}'\gamma_5}(p^{2},p'^{2},q^{2})\notag\\
&&p'_\mu\slashed {p}'\gamma_5+
\Pi^{\mathrm{QCD}}_{p'_\mu\slashed {p}'\slashed{p}\gamma_5}(p^{2},p'^{2},q^{2})~p'_\mu\slashed {p}'\slashed{p}\gamma_5+
\Pi^{\mathrm{QCD}}_{\slashed {p}'\gamma_\mu\gamma_5}(p^{2},p'^{2},q^{2})~\slashed {p}'\gamma_\mu\gamma_5+
\Pi^{\mathrm{QCD}}_{\slashed {p}'\gamma_\mu\slashed {p}\gamma_5}(p^{2},p'^{2},q^{2})~\slashed {p}'\gamma_\mu\slashed {p}\gamma_5+\Pi^{\mathrm{QCD}}_{p_{\mu} \slashed {p}' \slashed{p}\gamma_{5}}(p^{2},p'^{2},q^{2})\notag\\
&&p_{\mu} \slashed {p}' \slashed{p}\gamma_{5}+
 \Pi^{\mathrm{QCD}}_{\slashed{p}' \gamma_{\mu}}(p^{2},p'^{2},q^{2})~\slashed{p}' \gamma_{\mu}+
\Pi^{\mathrm{QCD}}_{p_\mu\slashed {p}'\gamma_5}(p^{2},p'^{2},q^{2})~p_\mu\slashed {p}'\gamma_5+
\Pi^{\mathrm{QCD}}_{p'_\mu\slashed {p}'}(p^{2},p'^{2},q^{2})~p'_\mu\slashed {p}'+
\Pi^{\mathrm{QCD}}_{p_\mu\slashed {p}'}(p^{2},p'^{2},q^{2})~p_\mu\slashed {p}'+\notag\\
&&\Pi^{\mathrm{QCD}}_{\gamma_\mu\slashed {p}\gamma_5}(p^{2},p'^{2},q^{2})~\gamma_\mu\slashed {p}\gamma_5+
\Pi^{\mathrm{QCD}}_{\gamma_{\mu}}(p^{2},p'^{2},q^{2})~\gamma_{\mu}+
\Pi^{\mathrm{QCD}}_{\gamma_{\mu}\slashed {p}}(p^{2},p'^{2},q^{2})~\gamma_{\mu}\slashed {p}+
\Pi^{\mathrm{QCD}}_{ \gamma_{\mu}\gamma_{5}}(p^{2},p'^{2},q^{2}) ~\gamma_{\mu}\gamma_{5}+\Pi^{\mathrm{QCD}}_{p_\mu\slashed {p}\gamma_5}(p^{2},p'^{2},q^{2})\notag\\
&&p_\mu\slashed {p}\gamma_5+
\Pi^{\mathrm{QCD}}_{p'_\mu\slashed {p}\gamma_5}(p^{2},p'^{2},q^{2})~p'_\mu\slashed {p}\gamma_5+
\Pi^{\mathrm{QCD}}_{p'_\mu\slashed {p}}(p^{2},p'^{2},q^{2})~p'_\mu\slashed {p}+
\Pi^{\mathrm{QCD}}_{p_\mu\slashed {p}}(p^{2},p'^{2},q^{2})~p_\mu\slashed {p}+\Pi^{\mathrm{QCD}}_{p'_\mu}(p^{2},p'^{2},q^{2})~p'_\mu+\notag\\
&&
\Pi^{\mathrm{QCD}}_{p'_\mu\gamma_5}(p^{2},p'^{2},q^{2})~p'_\mu\gamma_5+
\Pi^{\mathrm{QCD}}_{p_\mu}(p^{2},p'^{2},q^{2})~p_\mu+
\Pi^{\mathrm{QCD}}_{p_\mu\gamma_5}(p^{2},p'^{2},q^{2})~p_\mu\gamma_5.
\end{eqnarray}
The functions $ \Pi_i^{QCD}(p^2, p'^2, q^2)$ obtained in this part are Lorentz invariants and are rewritten in the form of double dispersion integrals:
\begin{eqnarray}\label{PiQCD}
\Pi^{\mathrm{QCD}}_i(p^{2},p'^{2},q^{2})&=&\int_{(m_u+m_d+m_s)^2}^{\infty}ds
\int_{(2m_u+m_d)^2}^{\infty}ds'~\frac{\rho
^{\mathrm{OPE}}_i(s,s',q^{2})}{(s-p^{2})(s'-p'^{2})}+\Gamma_i(p^2,p'^2,q^2).
\end{eqnarray}
The spectral density is defined as $\rho_i^{\mathrm{OPE}}(s,s',q^{2})=\frac{1}{\pi}Im\Pi^{QCD}_i(p^2,p'^2,q^2)$. In the calculations, the perturbative part as well as  the three-mass dimension $\langle \overline{q}q\rangle$ and the four-mass dimension $\langle G^2\rangle$ nonperturbative contributions as components of the spectral density   are included inside $  \rho_i^{\mathrm{OPE}}(s,s',q^{2})$. The contribution of the five-mass dimension nonperturbative operator  \(\langle \overline{q}g_{}\sigma Gq\rangle\)   denoted by the symbol $\Gamma_i$  in  Eq. (\ref{PiQCD}) is evaluated directly by applying the Borel transformation and not from the imaginary part of the correlation function.  As an example, the  expressions for the perturbative and nonperturbative contributions corresponding to the structure  $ \slashed {p'} \gamma_\mu \gamma_5$ are presented in the Appendix.  These contribute  in  the calculation of the form factor $G_1$.  By accepting the quark-hadron duality assumption, the continuum thresholds appear as $s_0$  and $s'_0$  for the initial and final states, respectively. Finally, by applying the double Borel transformation to remove the contributions of the excited and continuum states, the final form of the correlation function on the QCD side is obtained:
\begin{eqnarray}\label{qcd part2}
	\Pi^{\mathrm{QCD}}_i (M^2,M'^2,s_0,s'_0,q^2)=\int _{(m_u+m_d+m_s)^2}^{s_0} ds\int _{(2m_u+m_d)^2}^{s'_0}ds' e^{-s/M^2} e^{-s'/M'^2}\rho
	^{\mathrm{QCD}}_{i}(s,s',q^{2})+\tilde{\Gamma}_i(M^2,M'^2,q^2),
\end{eqnarray}
where $\tilde{\Gamma}_i$ is the double Borel-transformed form of $\Gamma_i$. Finally, by comparing the corresponding coefficients of the same Lorentz structures from the hadronic and QCD sides, six form factors are extracted.

\section { Numerical Analyses }\label{sec:three}
In this section, the numerical analysis of the form factors that represent the semileptonic decay process  $ \Lambda \rightarrow p  ~{\ell}\bar\nu_{\ell}$ is presented. The input parameters used in the calculations are listed in Table \ref{inputParameter}. In addition to these values, a set of auxiliary parameters is also required throughout the computations. These parameters include  the arbitrary parameter $\beta$, the two Borel parameters $(M^2, M'^2)$, and the continuum thresholds $(s_0, s'_0)$.  From the physical perspective and standard prescriptions of the method used, the final results are expected to show minimal sensitivity to variations in these parameters.
\begin{table}[h!]
		\begin{tabular}{|c|c|}
		\hline 
		Parameters                                             &  Values  \\
		\hline \hline
		$ m_u$                                                 & $(2.16^{+0.04}_{-0.04})~ \mathrm{MeV}$ \cite{ParticleDataGroup:2024cfk}\\
		$ m_d$                                                 & $(4.70^{+0.04}_{-0.04})~ \mathrm{MeV}$ \cite{ParticleDataGroup:2024cfk}\\
		$ m_s$                                                 & $(93.5^{+0.5}_{-0.5})~ \mathrm{MeV}$ \cite{ParticleDataGroup:2024cfk}\\
			$ m_e $                                                & $ 0.51~\mathrm{MeV}$ \cite{ParticleDataGroup:2024cfk}\\
		$ m_\mu $                                              & $ 105~\mathrm{MeV}$ \cite{ParticleDataGroup:2024cfk}\\
		$ m_{\Lambda}$                                       & $ (1115.68{}^{+0.006}_{-0.006}) ~\mathrm{MeV}$  \cite{ParticleDataGroup:2024cfk}\\
		$ m_p $                                      & $938.27~\mathrm{MeV}$  \cite{ParticleDataGroup:2024cfk} \\
		$ G_{F} $                                              & $ 1.17\times 10^{-5}~ \mathrm{GeV^{-2}}$ \cite{ParticleDataGroup:2024cfk}\\
		$ V_{ub} $                                             & $ (3.82\pm0.20)\times 10^{-3} $  \cite{ParticleDataGroup:2024cfk}\\
		$ m^2_0 $                                              & $ (0.8\pm0.2)~ \mathrm{GeV^2}$ \cite{Belyaev:1982sa,Belyaev:1982cd,Ioffe:2005ym} \\
		$\langle \bar{u} u\rangle$         & $-(0.24\pm0.01)^3 ~\mathrm{GeV^3}$  \cite{Belyaev:1982sa,Belyaev:1982cd} \\
		$\langle \bar{d} d\rangle$         & $-(0.24\pm0.01)^3 ~\mathrm{GeV^3}$  \cite{Belyaev:1982sa,Belyaev:1982cd} \\
		$\langle \bar{s} s\rangle$         & $-0.8(0.24\pm0.01)^3 ~\mathrm{GeV^3}$  \cite{Belyaev:1982sa,Belyaev:1982cd} \\
						$\langle \frac{\alpha_s}{\pi} G^2 \rangle $ & $(0.012\pm0.004)$ $~\mathrm{GeV}^4 $ \cite{Belyaev:1982sa,Belyaev:1982cd,Ioffe:2005ym}\\
				\hline 		
			\end{tabular}	
\caption{ Numerical inputs adopted in the analysis of the  decay   $ \Lambda \rightarrow p  ~{\ell}\bar\nu_{\ell}$.}\label{inputParameter}
\end{table}

 To determine the reliable range of these parameters within the framework of QCD sum rules, two essential criteria are considered: first, the dominance of the pole contribution (PC) over the contributions from the excited and continuum states, and second, the convergence of the OPE expansion. These conditions are implemented through the following relations:
\begin{equation} \label{PC}
	PC=\frac{\Pi^{QCD}(M^2,M'^2,s_0,s'_0)}{\Pi^{QCD}(M^2,M'^2,{\infty},{\infty})}\geq0.5
\end{equation}
and	
\begin{equation} \label{PC2}
	R(M^2, M'^2)=\frac{\Pi^{^{QCD}(dim4+dim5)}(M^2,M'^2,s_0,s'_0)}{\Pi^{QCD}(M^2,M'^2,s_0,s'_0)}\leq0.05.
\end{equation}
The criterion for the pole contribution [introduced in Eq. (\ref{PC})] is a condition according to which the contribution from the ground state must be larger than the contribution from the continuous spectrum and higher resonances. This condition sets the upper limits for the choice of the Borel parameters. On the other hand, the convergence condition of the OPE expansion [given in Eq. (\ref{PC2})] determines the lower limits of these parameters. Based on these two rules, the allowed range for the Borel parameters is established and implemented in the calculations as follows:
\begin{eqnarray}
	&&1.7~\mathrm{GeV^2}\leq M^2 \leq 2.1~\mathrm{GeV^2} \notag\\
	\mbox{and} \notag\\
	&&1.0~\mathrm{GeV^2} \leq M'^2 \leq 1.5~\mathrm{GeV^2}.
\end{eqnarray}

\begin{figure}[h!]
	\includegraphics[totalheight=4.5cm,width=6.5cm]{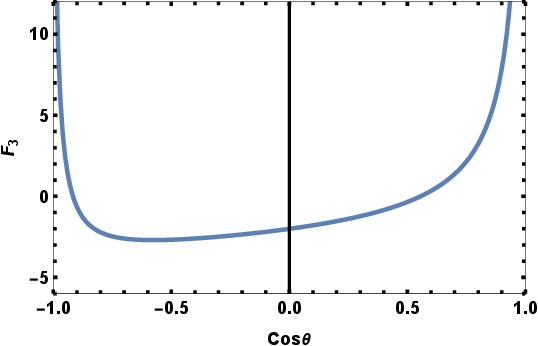}
	\caption{The dependence of the form factor $F_3$ associated with the structure $p'_{\mu}\slashed {p}$ on  $\cos\theta$ evaluated at the central values of   $s_0$, $s'_0$, $M^2$, $M'^2$ and at $q^2=0$.} \label{Fig:beta}
\end{figure}

The values of the continuum thresholds $(s_0 , s'_0)$ cannot be arbitrarily chosen, since these parameters depend on the excited states in each of the initial and final channels. Accordingly, suitable ranges for these thresholds are adopted to ensure the optimal stability of the sum rules: 
\begin{eqnarray}
	&&1.46~ \mathrm{GeV^2} \leq s_{0} \leq1.90~ \mathrm{GeV^2}\notag\\
	\mbox{and} \notag\\
	&&1.08~\mathrm{GeV^2}\leq s'_{0} \leq 1.84~ \mathrm{GeV^2}.
\end{eqnarray}
The last auxiliary parameter that needs to be determined is denoted by $\beta$, which appears in the calculations through Eqs. (\ref{cur1}) and (\ref{eq:CorrF2}). In principle, different values of $\beta$ can be chosen for the interpolating currents in the initial and final channels; however, our analysis indicates that their valid regions exhibit a significant overlap. Therefore, adopting a common value of $\beta$ for both channels is reasonable and optimal. To determine the range of this parameter, we note that $\beta$ can extend over the entire interval from $-\infty$ to $+\infty$. Therefore, it is redefined as $\beta=\tan\theta$ so that all possible values are covered through a new variable. In this representation, the domain of $\cos\theta$ is restricted to the interval $[-1, +1]$. Figure \ref{Fig:beta}, as an example,  illustrates the behavior of the  form factor $F_3$ with respect to variations in $\cos\theta$, thereby allowing the identification of the suitable region for this variable. Based on numerical calculations and the patterns extracted from the figure  and  figures for other form factors,  the region that ensures the stability (minimal variation)  of the form factors under variations of $\cos\theta$ can be identified commonly for all form factors as follows:
\begin{eqnarray}\label{beta fun}
-0.72 \le\ \cos\theta \le\ -0.57.
\end{eqnarray}

\begin{figure}[h!] 
	\includegraphics[totalheight=4.8cm,width=4.9cm]{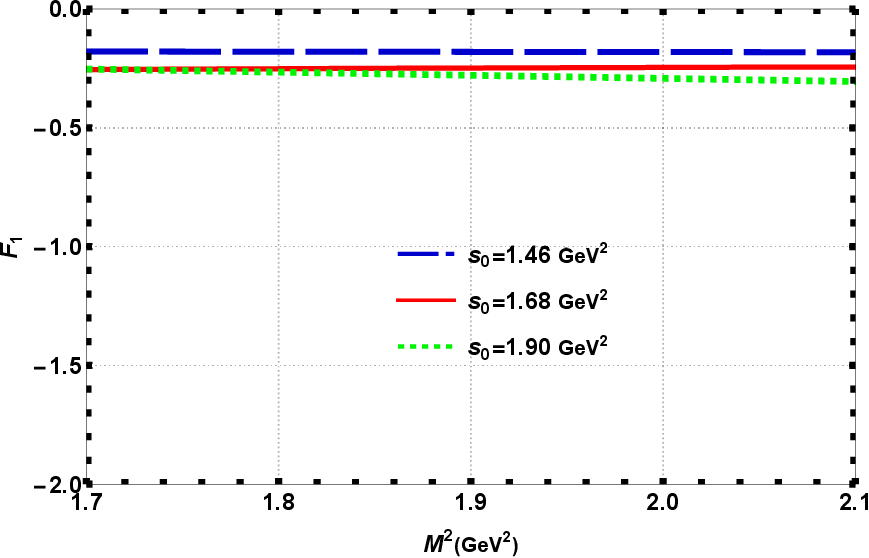}
	\includegraphics[totalheight=4.8cm,width=4.9cm]{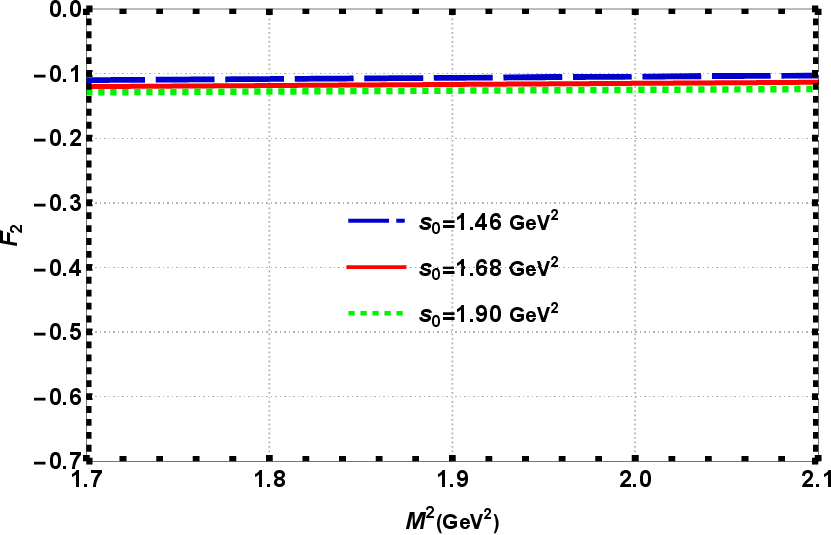}
	\includegraphics[totalheight=4.8cm,width=4.9cm]{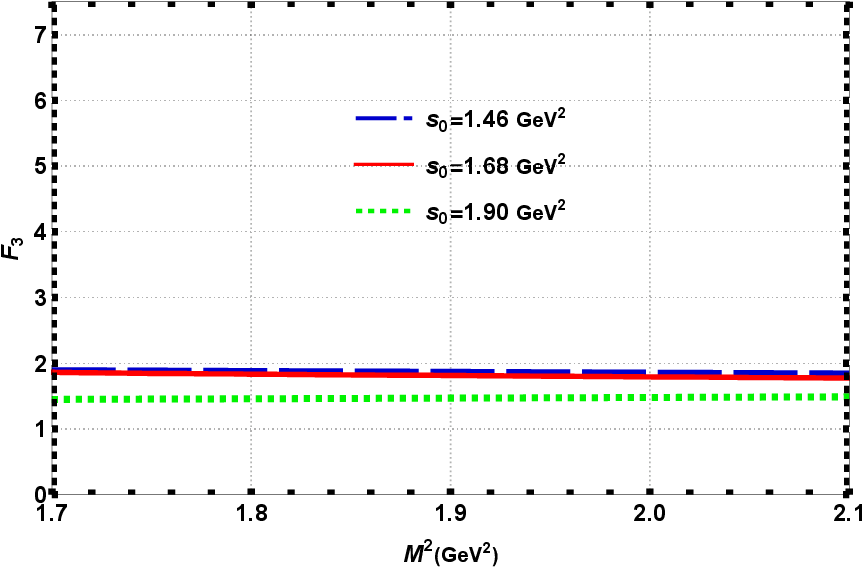}
	\includegraphics[totalheight=4.8cm,width=4.9cm]{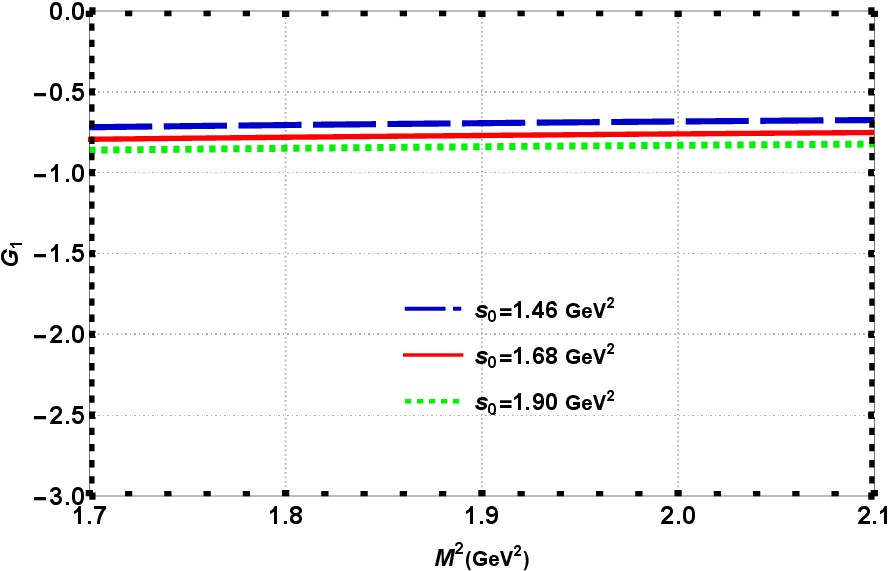}
	\includegraphics[totalheight=4.8cm,width=4.9cm]{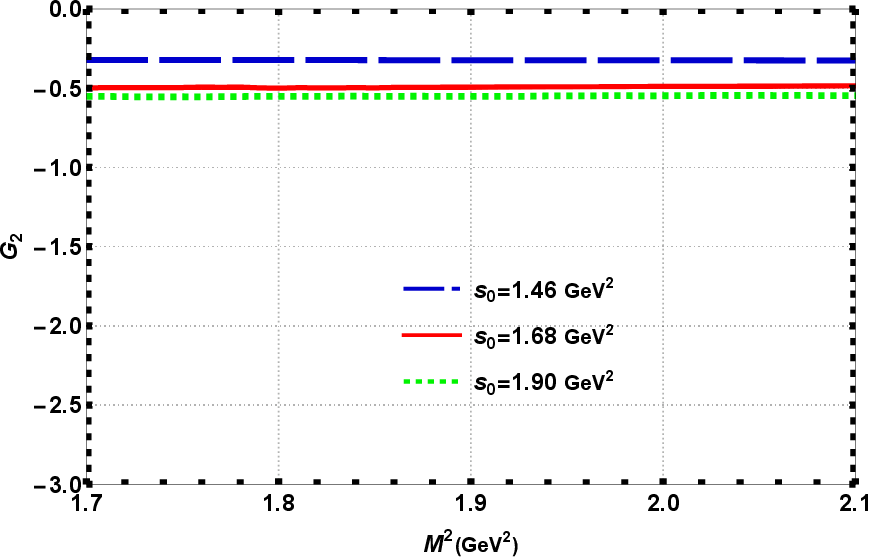}
	\includegraphics[totalheight=4.8cm,width=4.9cm]{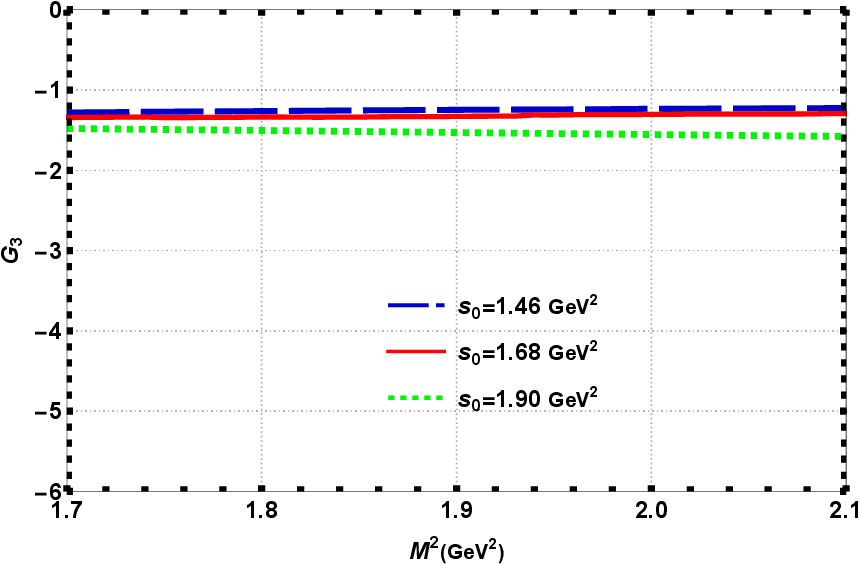}
	\caption{Dependence of the form factors on the Borel parameter $M^2$ for various $s_0$, evaluated at $q^2=0$ and the mean values of other auxiliary parameters.}\label{Fig:BorelM}
\end{figure}

\begin{figure}[h!]
	\includegraphics[totalheight=4.8cm,width=4.9cm]{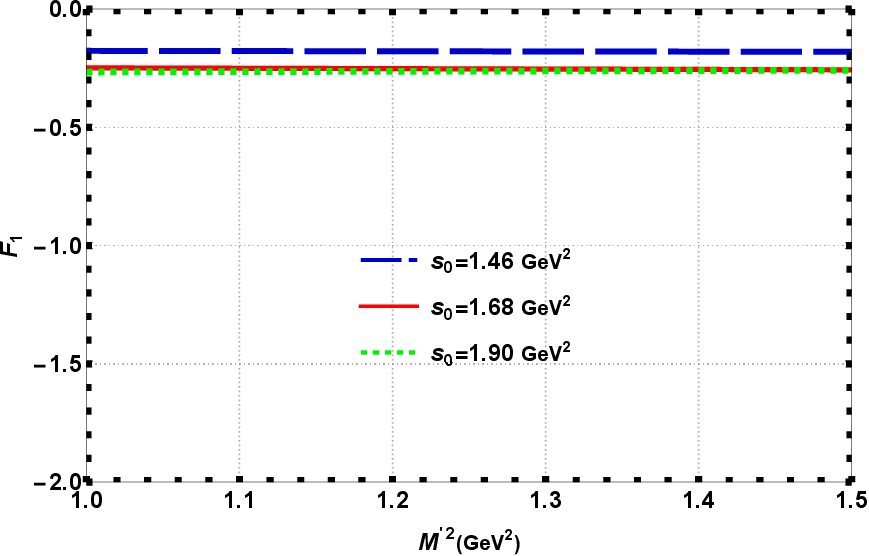}
	\includegraphics[totalheight=4.8cm,width=4.9cm]{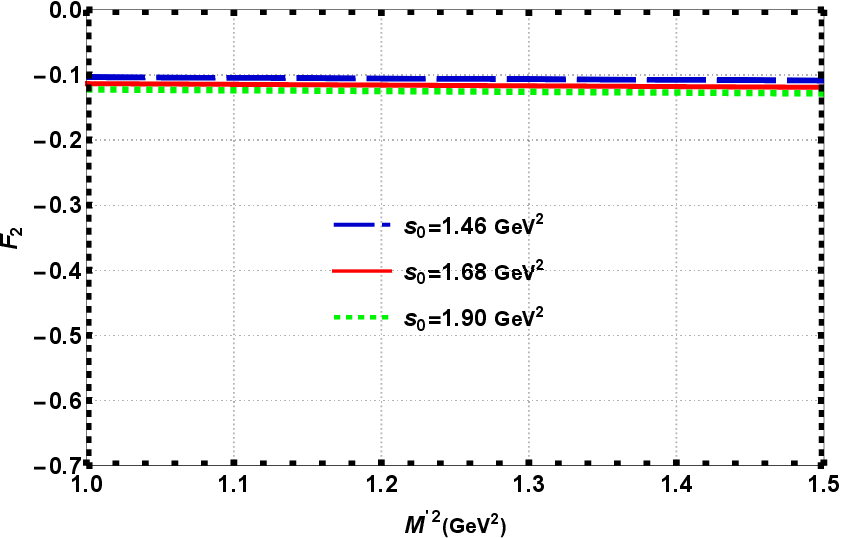}
	\includegraphics[totalheight=4.8cm,width=4.9cm]{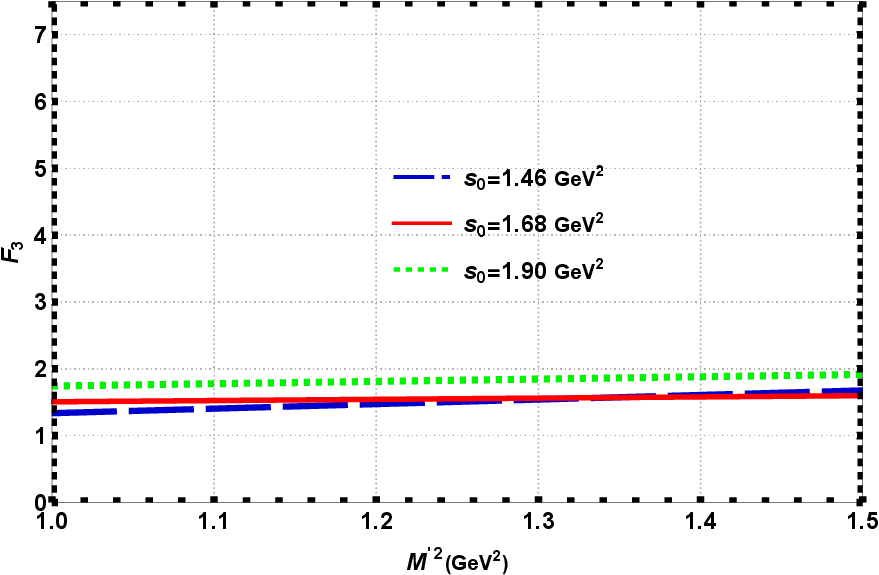}
	\includegraphics[totalheight=4.8cm,width=4.9cm]{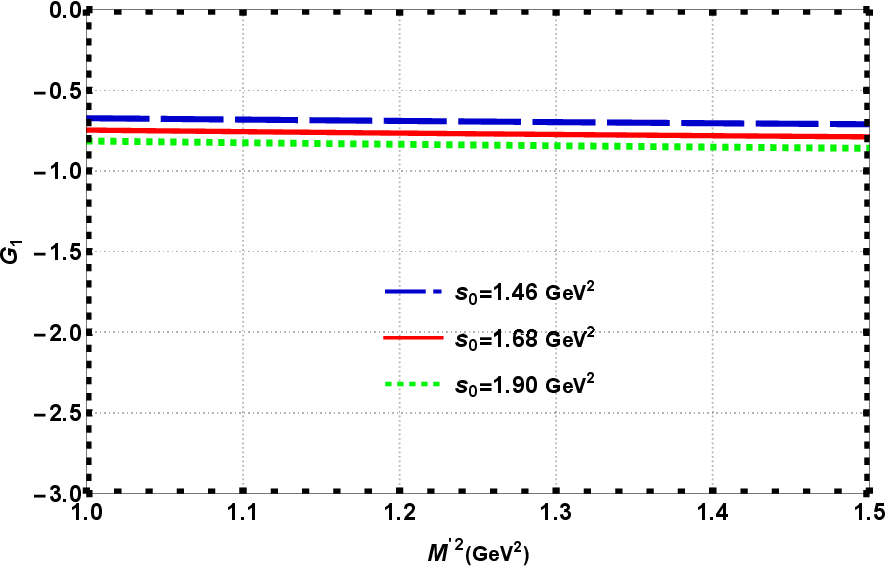}
	\includegraphics[totalheight=4.8cm,width=4.9cm]{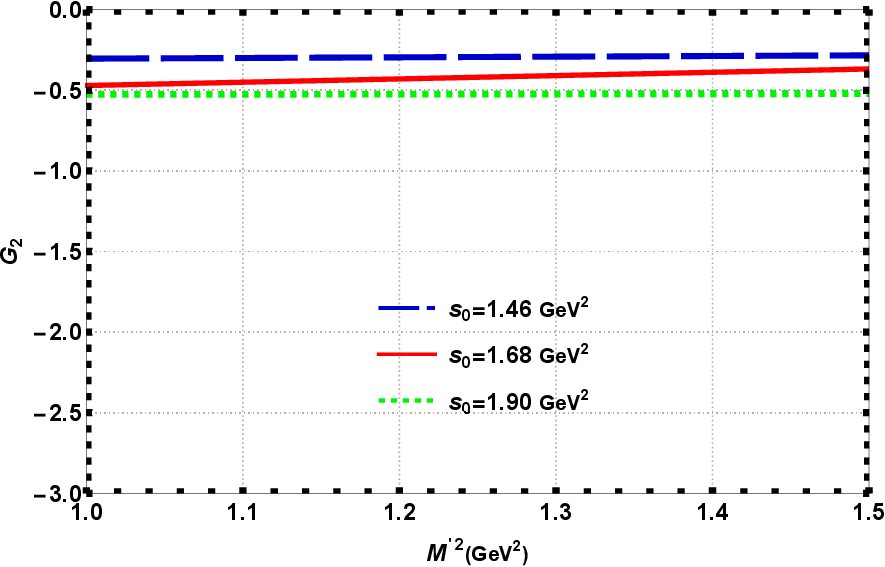}
	\includegraphics[totalheight=4.8cm,width=4.9cm]{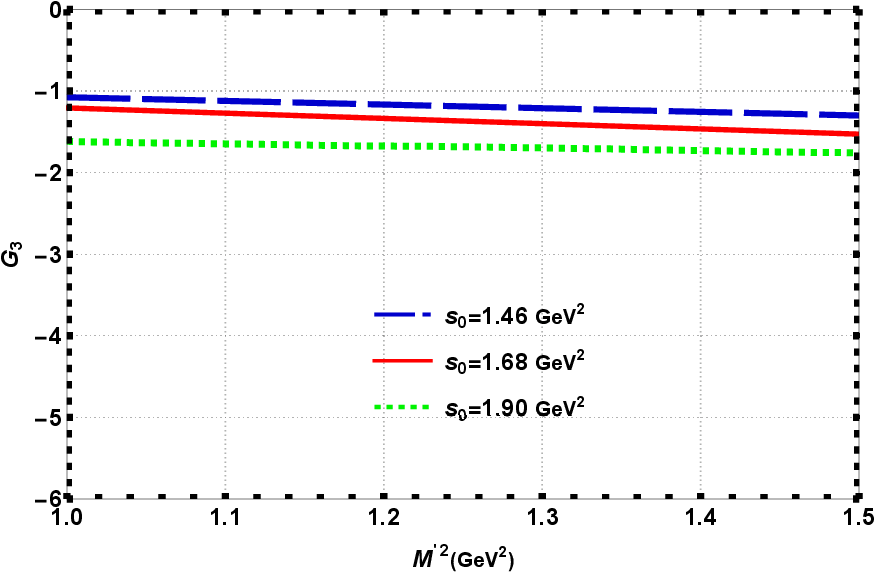}
	\caption{Dependence of the form factors on the Borel parameter $M'^2$ for various $s_0$, evaluated at $q^2=0$ and the mean values of other auxiliary parameters.} \label{Fig:BorelMM}
\end{figure}

\begin{figure}[h!] 
	\includegraphics[totalheight=4.8cm,width=4.9cm]{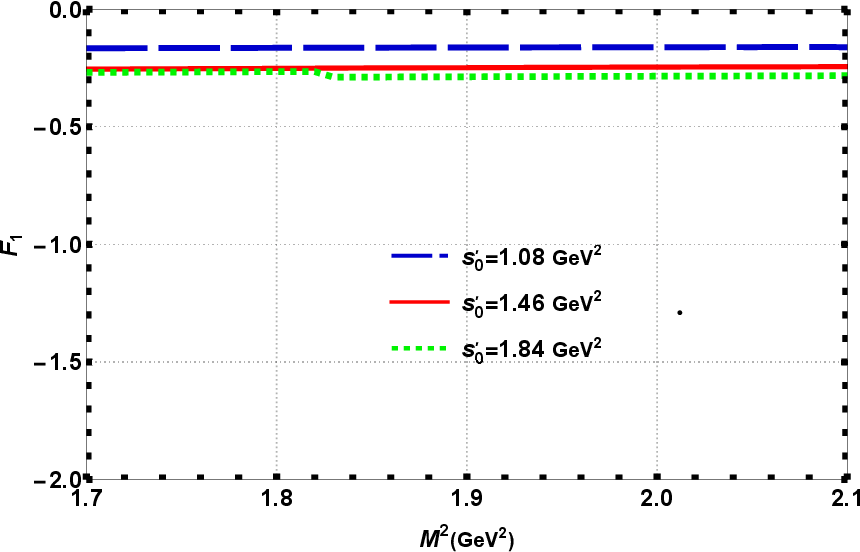}
	\includegraphics[totalheight=4.8cm,width=4.9cm]{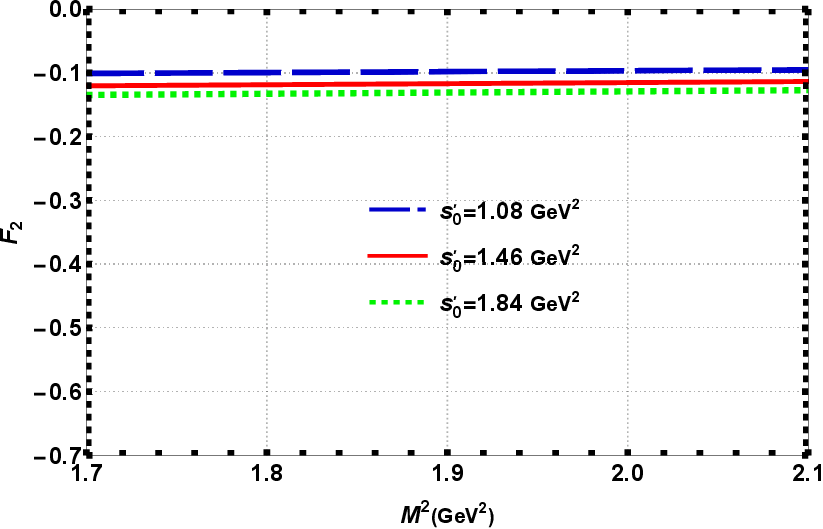}
	\includegraphics[totalheight=4.8cm,width=4.9cm]{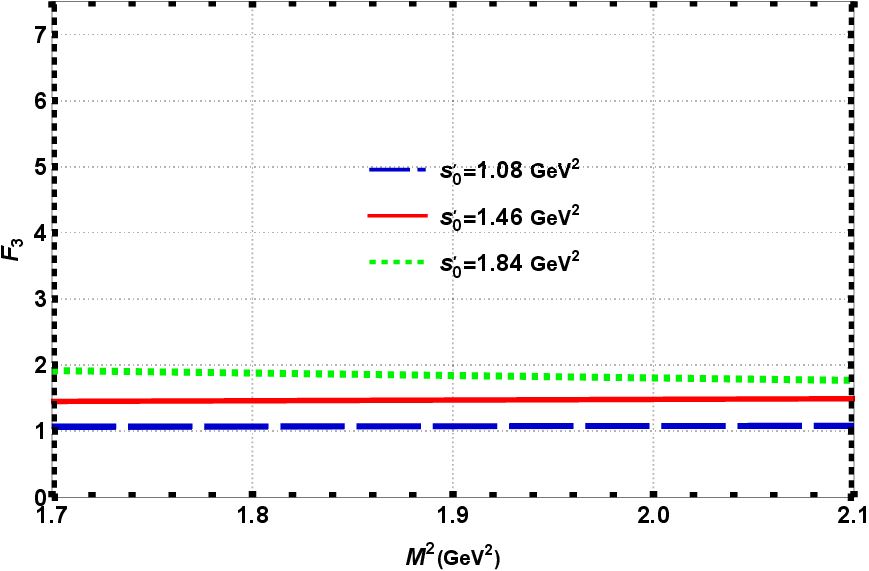}
	\includegraphics[totalheight=4.8cm,width=4.9cm]{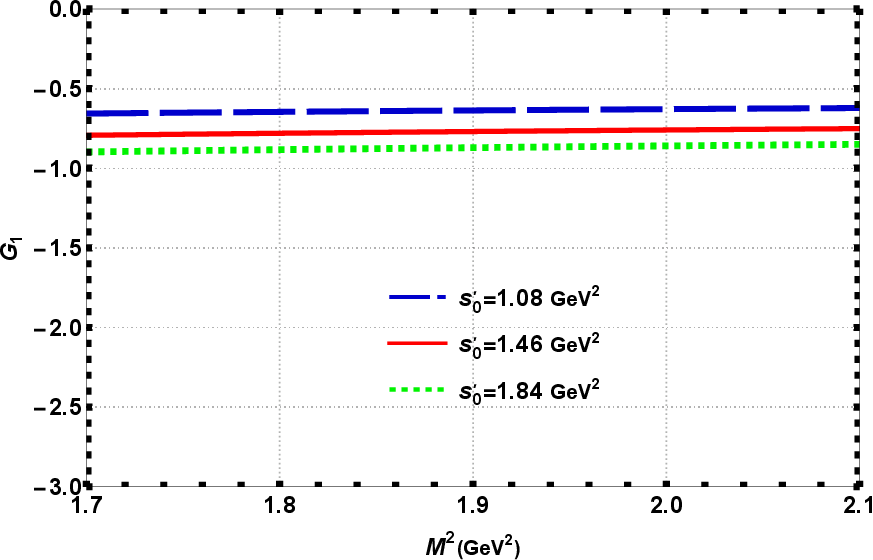}
	\includegraphics[totalheight=4.8cm,width=4.9cm]{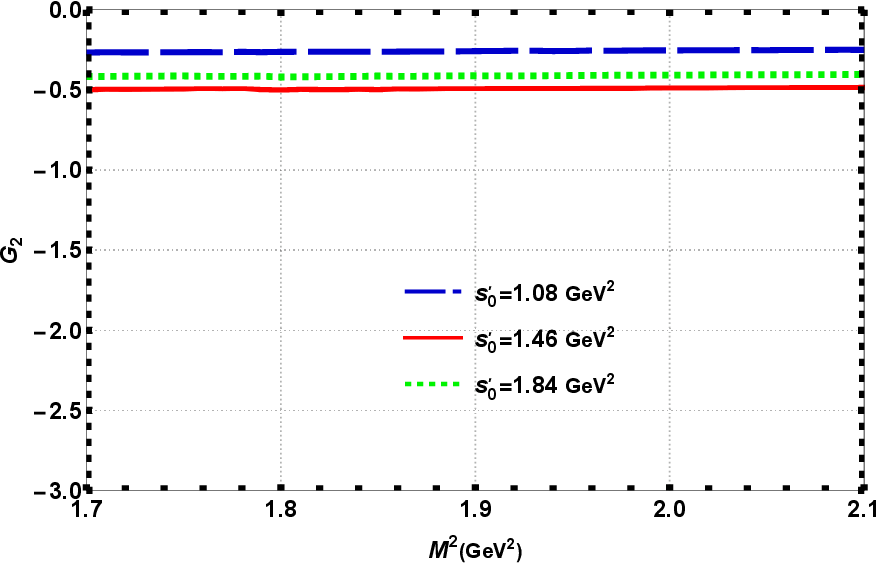}
	\includegraphics[totalheight=4.8cm,width=4.9cm]{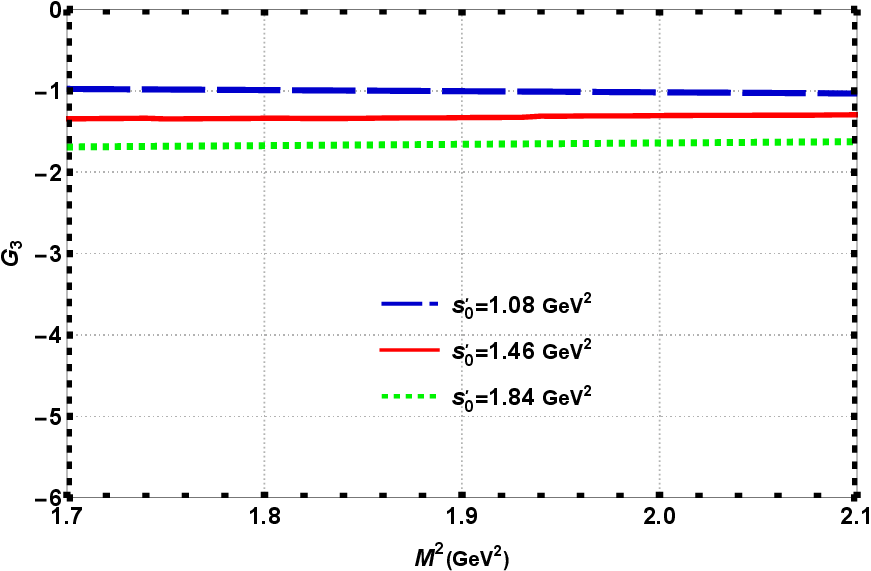}
	\caption{Dependence of the form factors on the Borel parameter $M^2$ for various $s'_0$, evaluated at $q^2=0$ and the mean values of other auxiliary parameters.}\label{Fig:BorelMs'}
\end{figure}

\begin{figure}[h!]
	\includegraphics[totalheight=4.8cm,width=4.9cm]{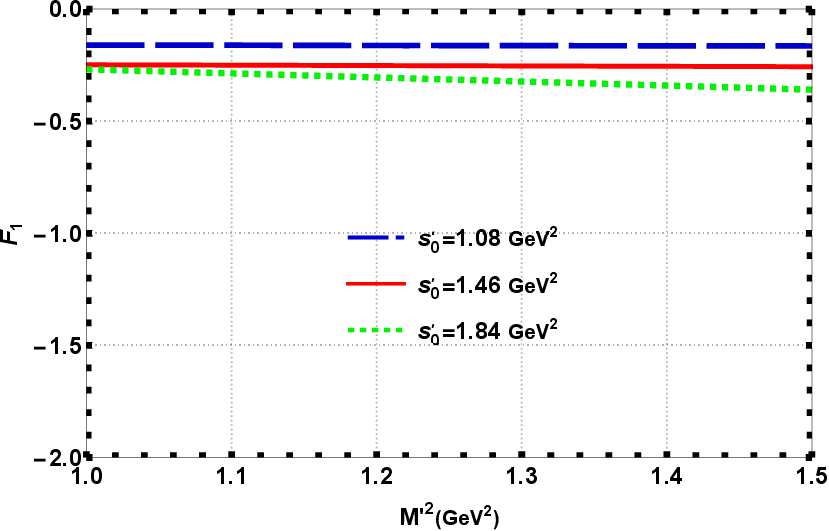}
	\includegraphics[totalheight=4.8cm,width=4.9cm]{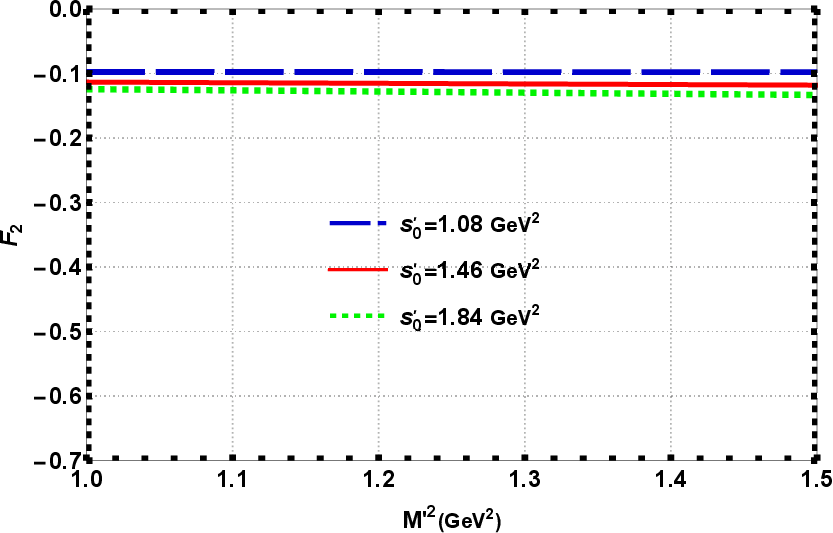}
	\includegraphics[totalheight=4.8cm,width=4.9cm]{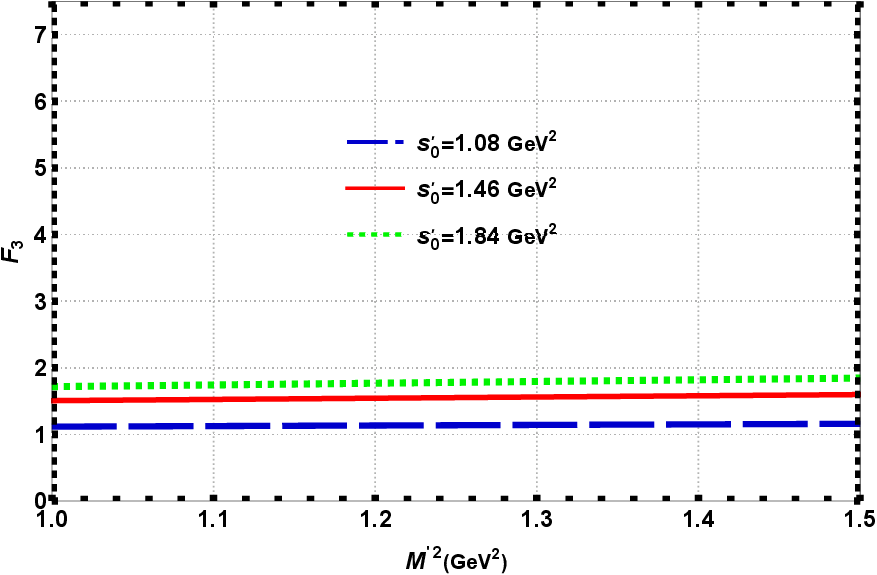}
	\includegraphics[totalheight=4.8cm,width=4.9cm]{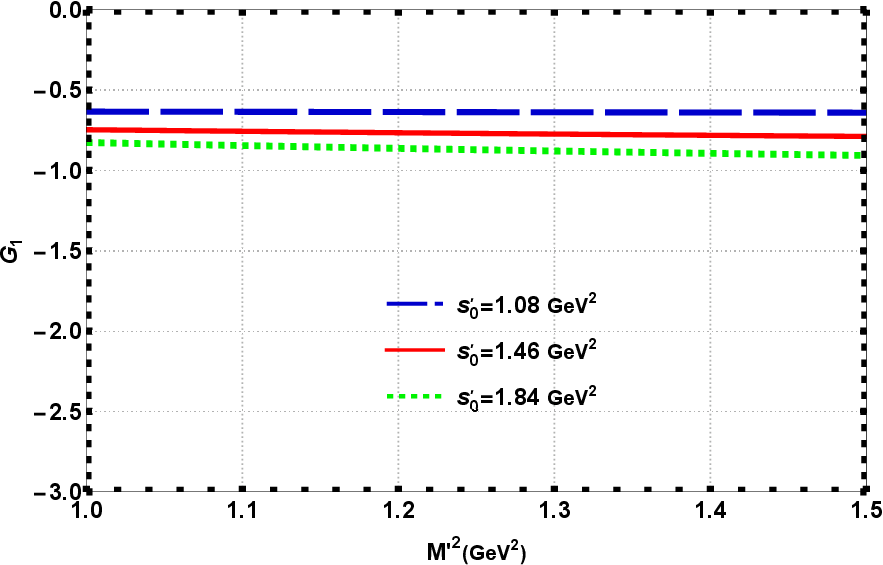}
	\includegraphics[totalheight=4.8cm,width=4.9cm]{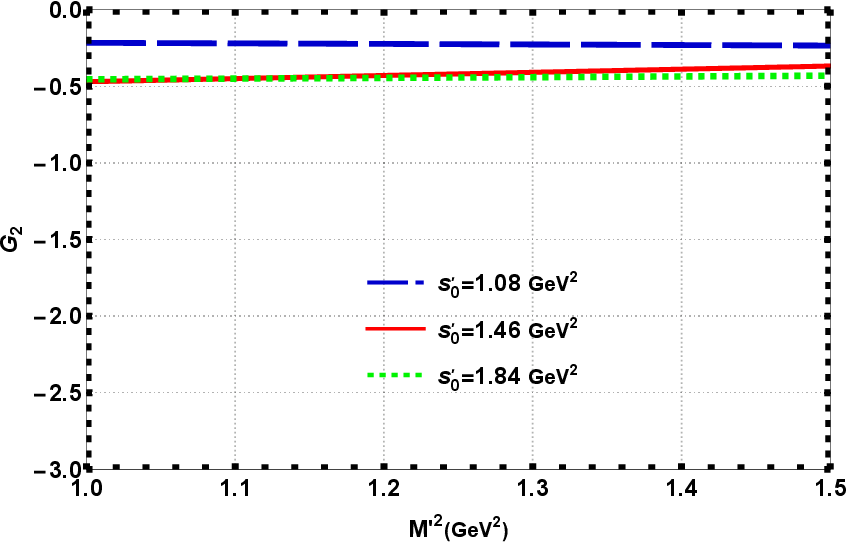}
	\includegraphics[totalheight=4.8cm,width=4.9cm]{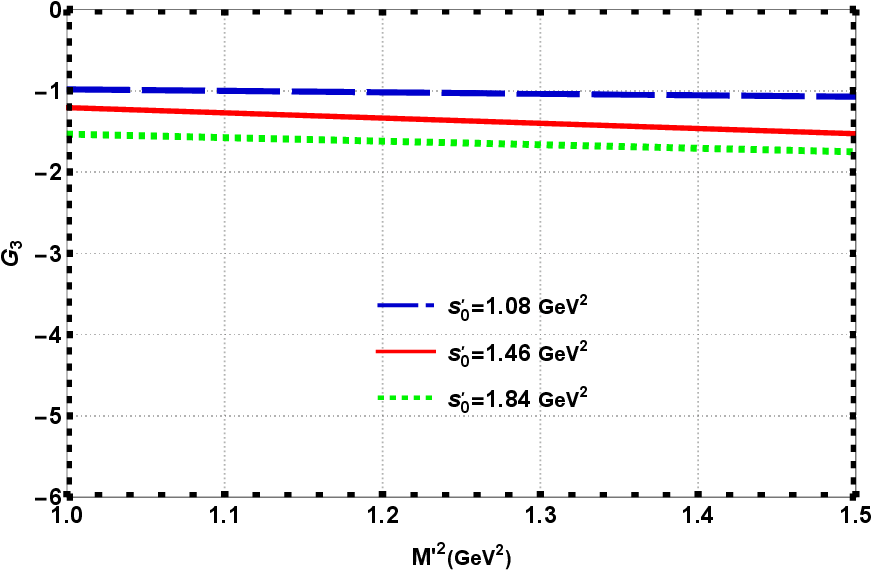}
	\caption{Dependence of the form factors on the Borel parameter $M'^2$ for various $s'_0$, evaluated at $q^2=0$ and the mean values of other auxiliary parameters.} \label{Fig:BorelMMs'}
\end{figure}

	To examine the sensitivity of the results with respect to the auxiliary parameters, Figs.~(\ref{Fig:BorelM}--\ref{Fig:BorelMMs'})  are plotted. In these figures, the six form factors $F_1$, $F_2$, $F_3$, $G_1$, $G_2$, and $G_3$,  which are associated with different Lorentz structures  $\slashed {p}'\gamma_{\mu}\slashed {p}$, $p_\mu \slashed {p}'$, $p'_{\mu}\slashed {p}$,  $ \slashed {p'} \gamma_\mu \gamma_5$, $p_\mu \slashed {p}' \gamma_{5}$ and  $p'_\mu \slashed {p} \gamma_{5}$, are examined.  In  Figs. \ref{Fig:BorelM} and  \ref{Fig:BorelMM}, the behavior of the form factors as functions of the Borel parameters $M^2$ and $M'^2$ is shown, while the continuum threshold $s_0$ is fixed at three specific values and all other parameters are set to their average values. In Figs. \ref{Fig:BorelMs'} and \ref{Fig:BorelMMs'}, the same analysis is carried out for three different values of $s'_0$, with the other parameters kept at their average values. All figures are calculated at the point $q^2 = 0$. The results indicate that the form factors exhibit stable and reliable behavior within the defined ranges of the auxiliary parameters.
 To obtain a smooth and reliable behavior of the form factors across the entire physical region of the momentum transfer $q^2$, it is necessary to fit the numerical results derived from QCD sum rules in a limited domain to a suitable analytical function. This procedure is employed because the sum-rule method is valid only within a limited region of $q^2$, whereas the evaluation of the decay rate and other phenomenological quantities requires the form factors over the entire physical region $ m_l^2\leq q^2 \leq (m_{\Lambda}-m_{p})^2$. To ensure the robustness and model independence of our analysis, we employed two complementary parametrizations: the phenomenological polynomial fit and the theoretically motivated $z$ expansion. The polynomial fit provides a straightforward phenomenological benchmark and captures the local curvature of the sum-rule data near $q^2 = 0$ through a relatively large number of free parameters. However, this flexibility is purely mathematical and can lead to unphysical behavior at large $q^2$. In contrast, the $z$ expansion exploits the analytic structure of the form factor via a conformal mapping based on the lightest physical branch point, requiring fewer parameters while ensuring controlled convergence and stability. It maps the singularity to the boundary of the unit disk, guaranteeing that the expansion coefficients ($a_k$) describe only the physical region, avoiding unphysical artifacts. Together, these schemes provide a robust and reliable characterization of the form factor across the kinematic range. To this end, we have utilized the following two fitted functions:
 \begin{equation} \label{fitffunction1}
	{\cal F}(q^2)=\frac{{\cal
			F}(0)}{\displaystyle\left(1-a_1\frac{q^2}{m^2_{\Lambda}}+a_2
		(\frac{q^2}{m_{\Lambda}^2})^2+a_3(\frac{q^2}{m_{\Lambda}^2})^3+a_4(\frac{q^2}{m_{\Lambda}^2})^4\right)}
\end{equation} 
 and
 \begin{equation}\label{fitffunction2}
 F(q^2) = \frac{1}{1 - \dfrac{q^2}{m^2}}
         \left[ F_0 + A_1 \big( z(q^2) - z(0) \big) \right],
\end{equation}
 where $z(q^2)$ is defined as:
\begin{equation}
z(q^2) = \frac{\sqrt{t_+ - q^2} - \sqrt{t_+ - t_0}}
              {\sqrt{t_+ - q^2} + \sqrt{t_+ - t_0}},
\end{equation}
where $t_{\pm}=(m_{\Lambda}\pm m_p)^2$, $t_0=t_{+}(1-\sqrt{1-t_{-}/t_{+}})$, and the parameter $m$ corresponds to the pole mass, which ensures that the analytic behavior of the form factors near the poles close to the physical $q^2$ region is properly taken into account in the $z$ expansion. In the semileptonic decay under consideration, we set $m$ = 0.892~\text{GeV} for the vector form factors and $m$ = 1.27~\text{GeV} for the axial form factors, since for the underlying $s \to u$ transition the nearest resonances generating the poles are the $K^*$ and $K_1$, respectively \cite{Detmold:2015aaa}. The first fitted function [Eq. (\ref{fitffunction1})] will be referred to as the polynomial fit (in the denominator), while the second one [Eq. (\ref{fitffunction2})] corresponds to the $z$ expansion parametrization. In Eq. (\ref{fitffunction1}), the quantities ${\cal F}(0)$, $a_1$, $a_2$, $a_3$ and $a_4$, and in Eq. \ref{fitffunction2} the quantities $F_0$ and $A_1$  , are considered  fitting parameters for the structures introduced in Tables \ref{Tab:parameterfit1} and  \ref{Tab:parameterfit2}. 

\begin{table}[h!]
	\begin{ruledtabular}
		\begin{tabular}{|c|c|c|c|c|c|c|}
					& $F_1(q^2): \slashed {p}'\gamma_{\mu}\slashed {p}$ & $F_2(q^2):p_\mu \slashed {p}'$  & $F_3(q^2):p'_{\mu}\slashed {p}$   & $G_1(q^2):  \slashed {p'} \gamma_\mu \gamma_5$ & $G_2(q^2):p_\mu \slashed {p}' \gamma_{5}$  & $G_3(q^2):p'_\mu \slashed {p} \gamma_{5}$       \\
			\hline
			${\cal F}(q^2=0)$ & $-0.24\pm0.08$        & $-0.11\pm0.03$      & $1.58\pm0.47$     & $-0.76{}^{+0.21}_{-022}$  & $-0.44\pm0.13$  & $-1.50\pm0.45$ \\
			$a_1$           & $13.33$          & $19.23$            &$13.51$            & $1.85$           & $16.66$            &$ 9.96$            \\
			$a_2$           & $0.90$         & $-23.84$           &$-78.83$              & $129.32$          & $-24.31$           & $-27.91$           \\
			$a_3$           & $0.99$            & $268.50$           &$935.739$           & $-6565.64$          & $268.50$           & $459.20$          \\
			$a_4$           & $0.99$           & $-1174.79$           &$-3550.64$             & $-188511.33$         & $-1174.79$          & $-2722.99$           \\
		\end{tabular}
		\caption{Parameters for the polynomial fitting of all form factors related to $\Lambda \to p  l\bar\nu_{\ell}$.}\label{Tab:parameterfit1}
	\end{ruledtabular}
\end{table}

\begin{figure}[h!] 
	\includegraphics[totalheight=4.5cm,width=4.8cm]{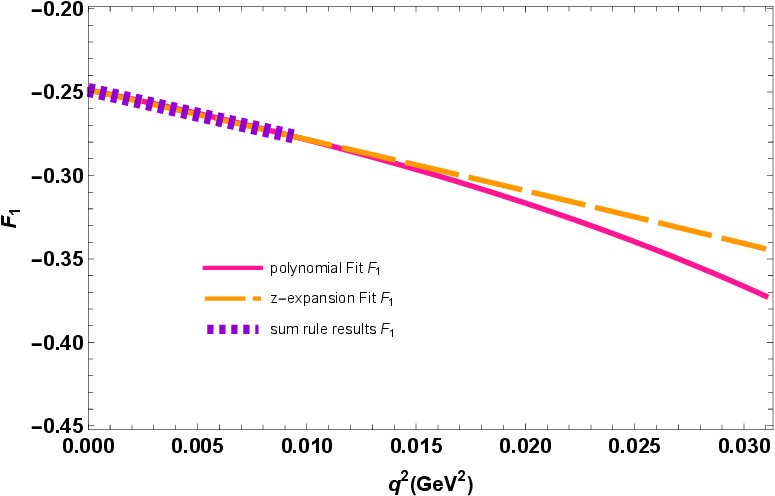}
	\includegraphics[totalheight=4.5cm,width=4.8cm]{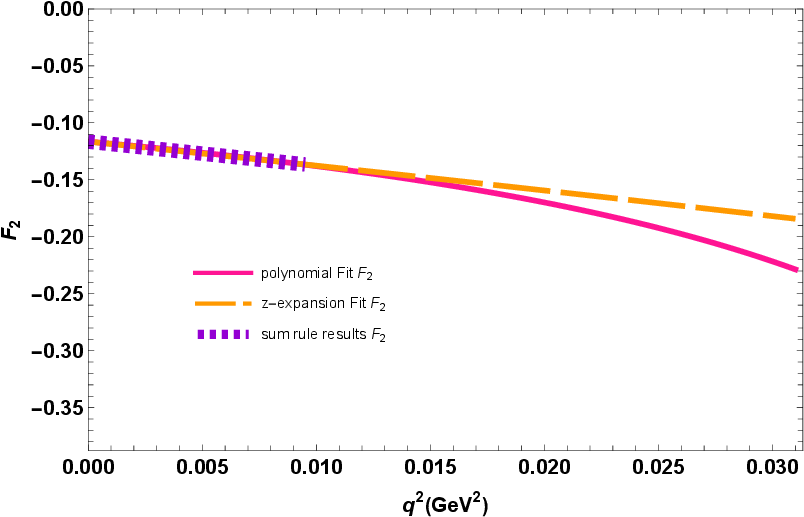}
	\includegraphics[totalheight=4.5cm,width=4.8cm]{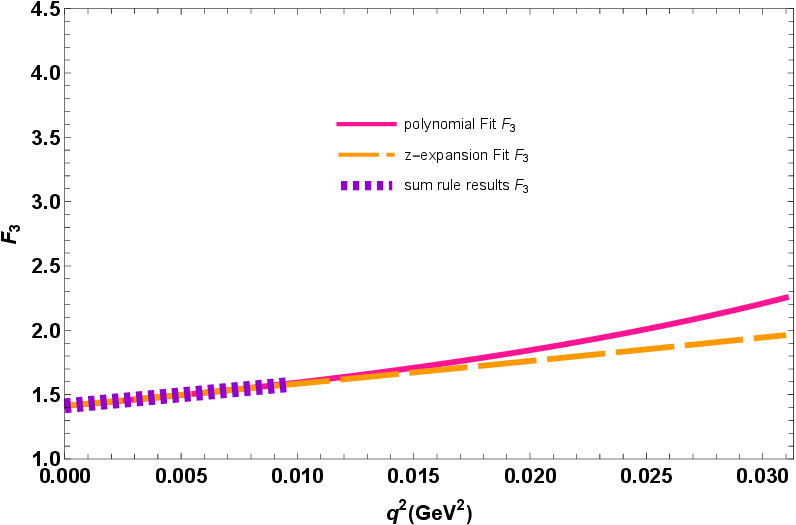}
	\includegraphics[totalheight=4.5cm,width=4.8cm]{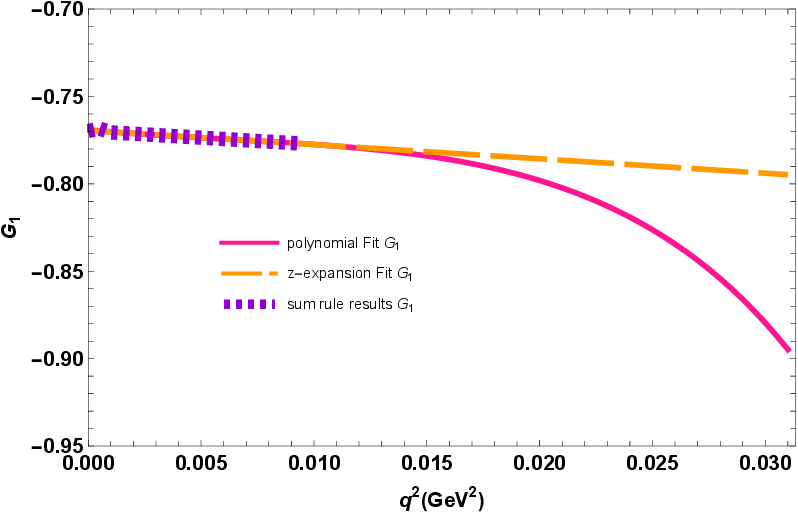}
	\includegraphics[totalheight=4.5cm,width=4.8cm]{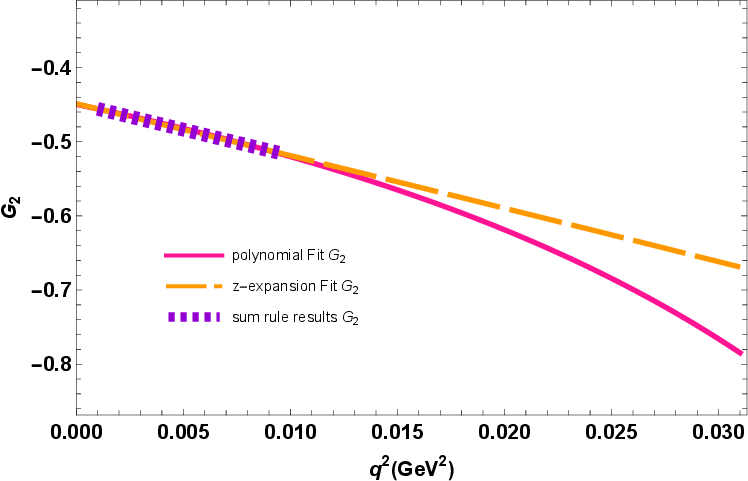}
	\includegraphics[totalheight=4.5cm,width=4.8cm]{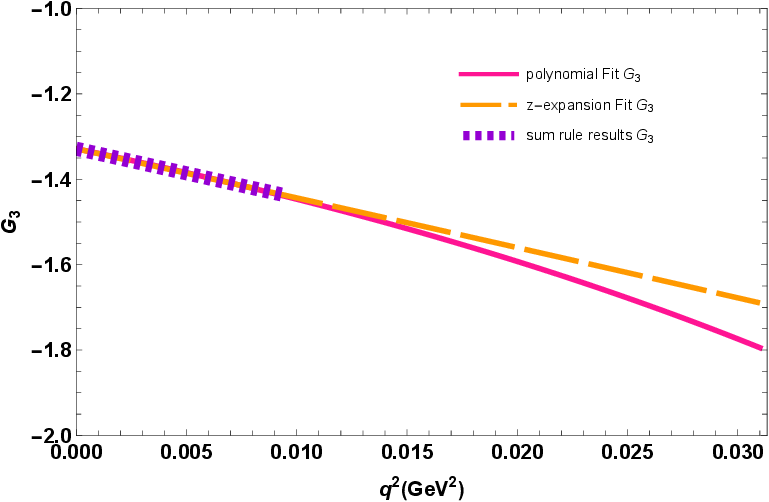}
	\caption{The form factors $F_1$, $F_2$, $F_3$, $G_1$, $G_2$, and $G_3$ associated with the structures in Tables \ref{Tab:parameterfit1} and \ref{Tab:parameterfit2}, as functions of $q^2$, plotted using polynomial fits, $z$ expansion, and on, and sum-rule results, with auxiliary parameters set at their central values.}\label{Fig:formfactor1}
\end{figure}

\begin{figure}[h!] 
	\includegraphics[totalheight=4.5cm,width=4.8cm]{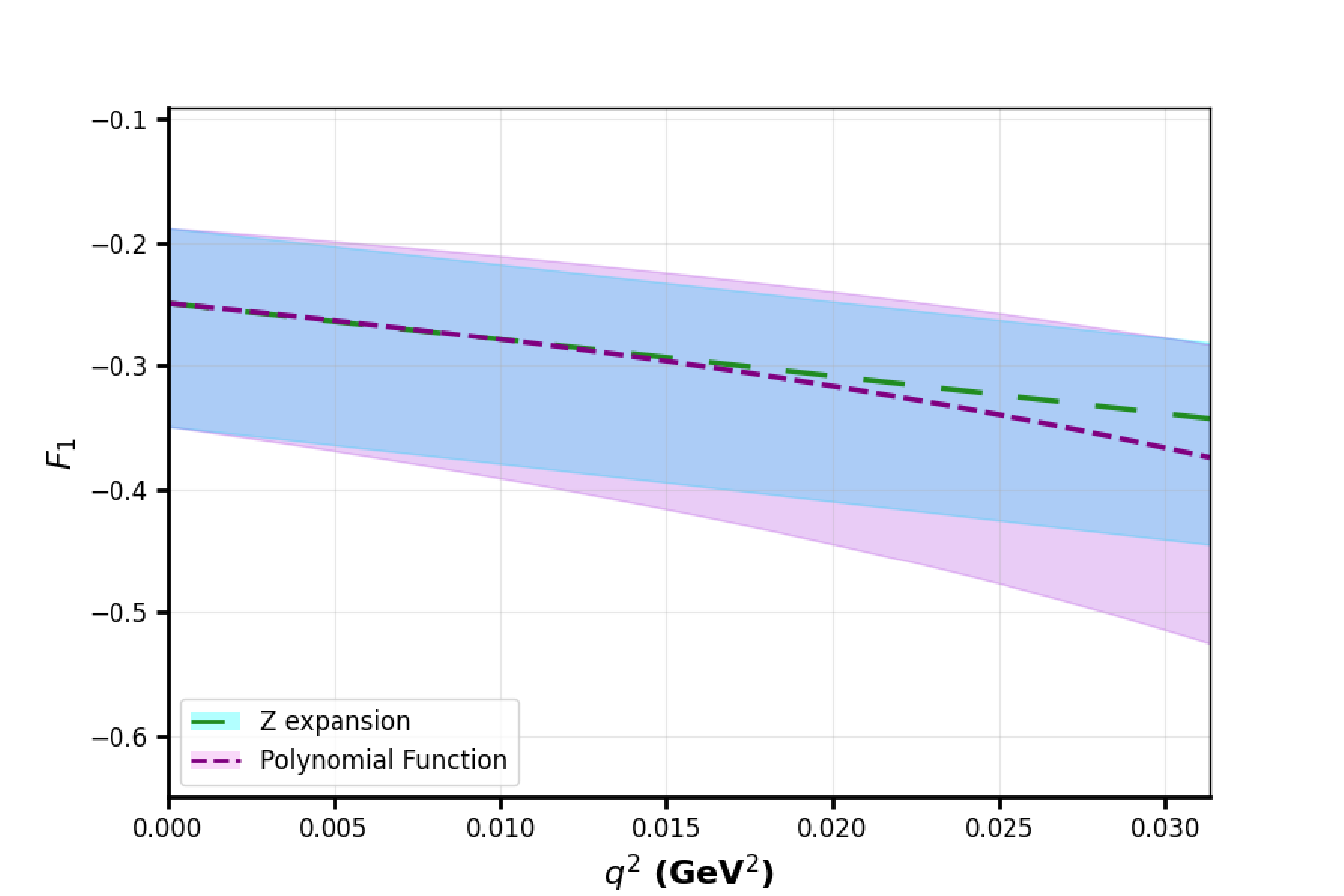}
	\includegraphics[totalheight=4.5cm,width=4.8cm]{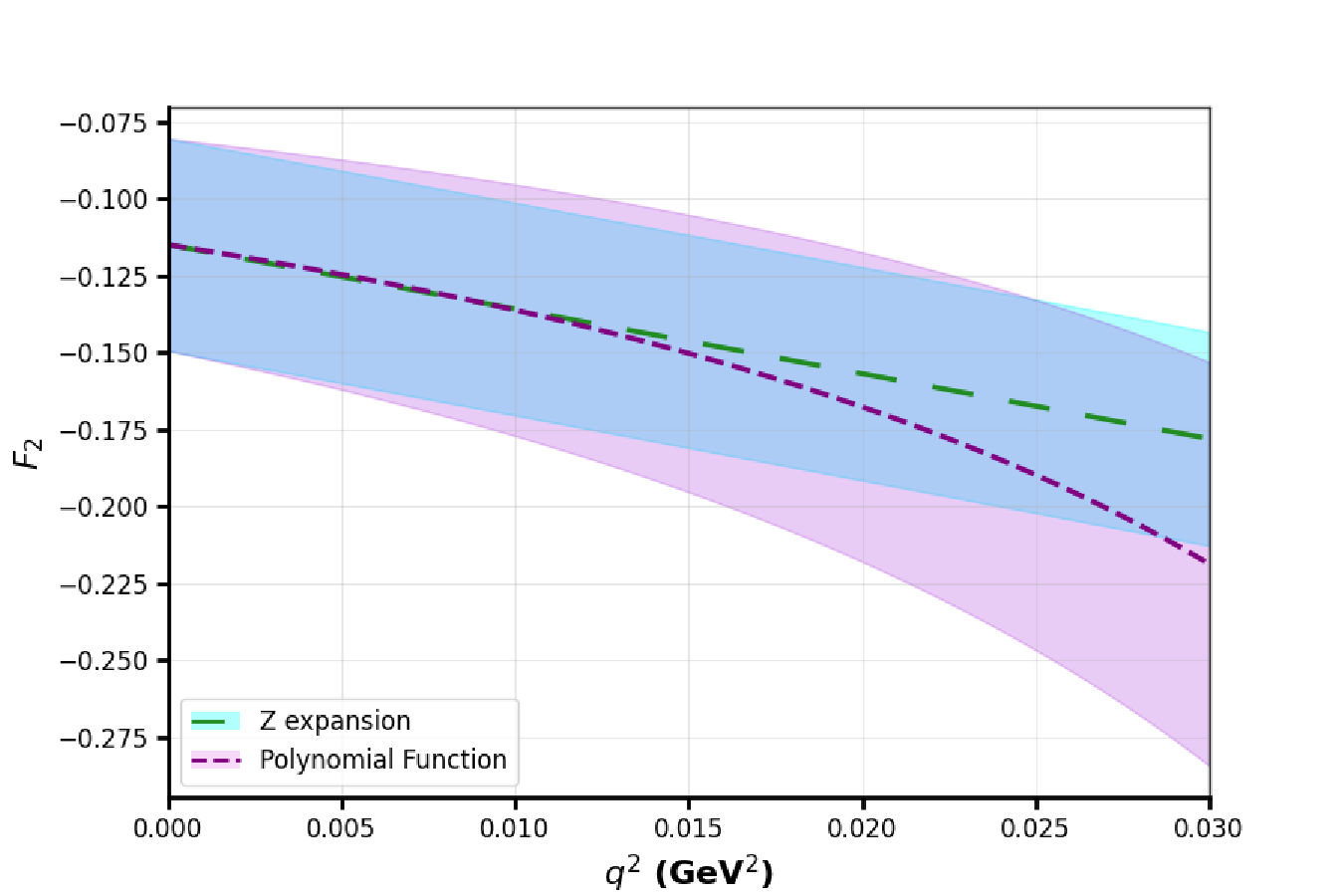}
	\includegraphics[totalheight=4.5cm,width=4.8cm]{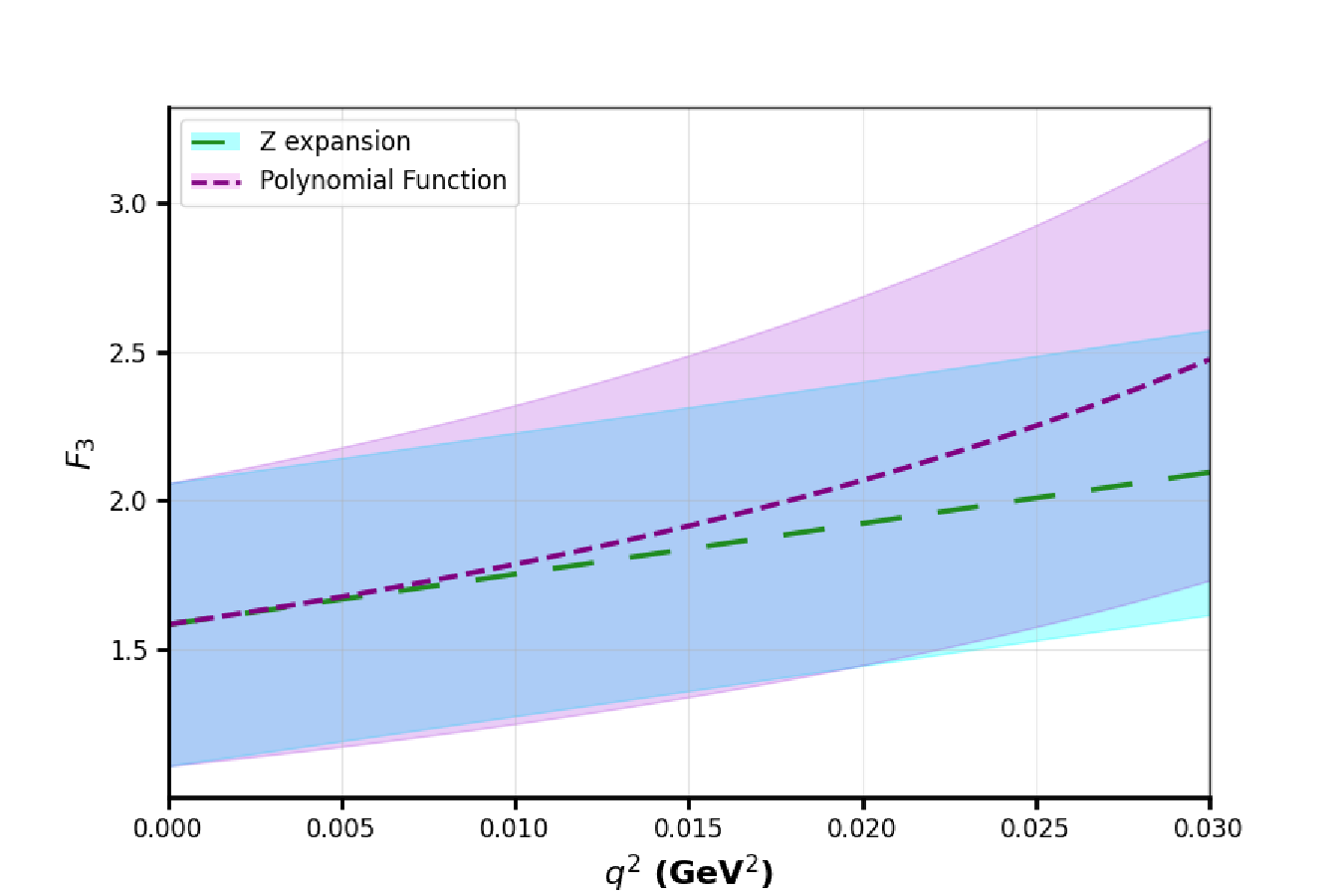}
	\includegraphics[totalheight=4.5cm,width=4.8cm]{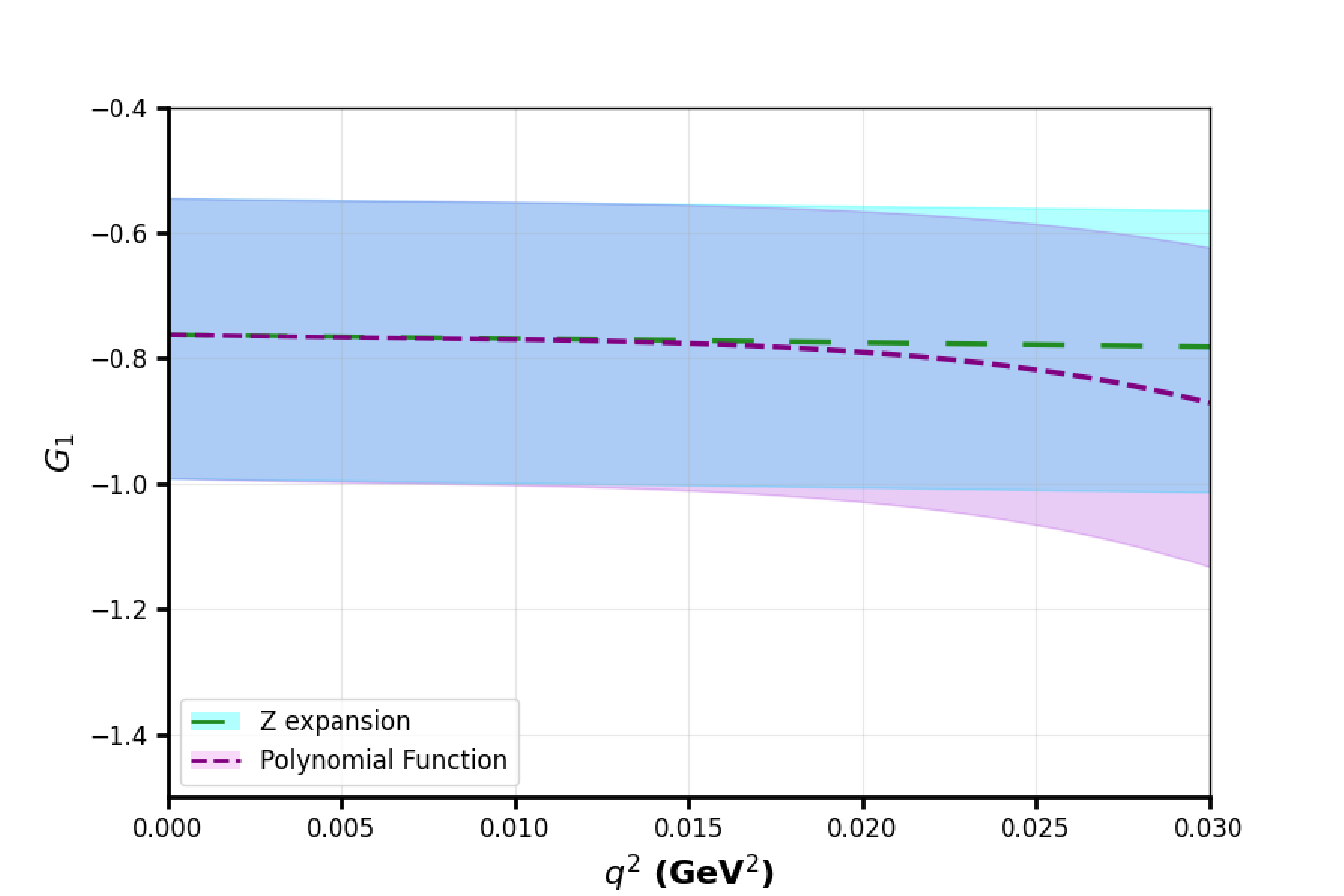}
	\includegraphics[totalheight=4.5cm,width=4.8cm]{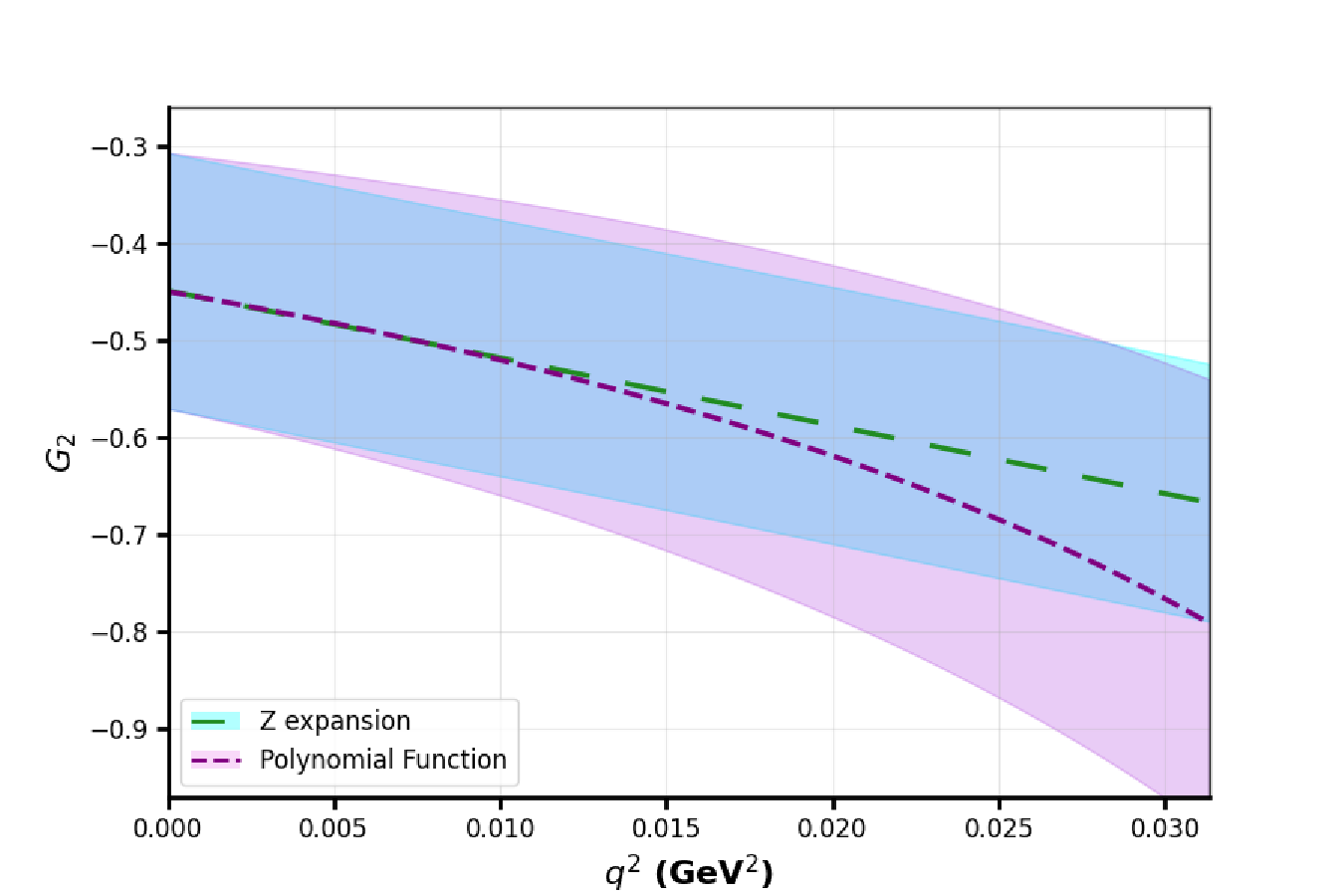}
	\includegraphics[totalheight=4.5cm,width=4.8cm]{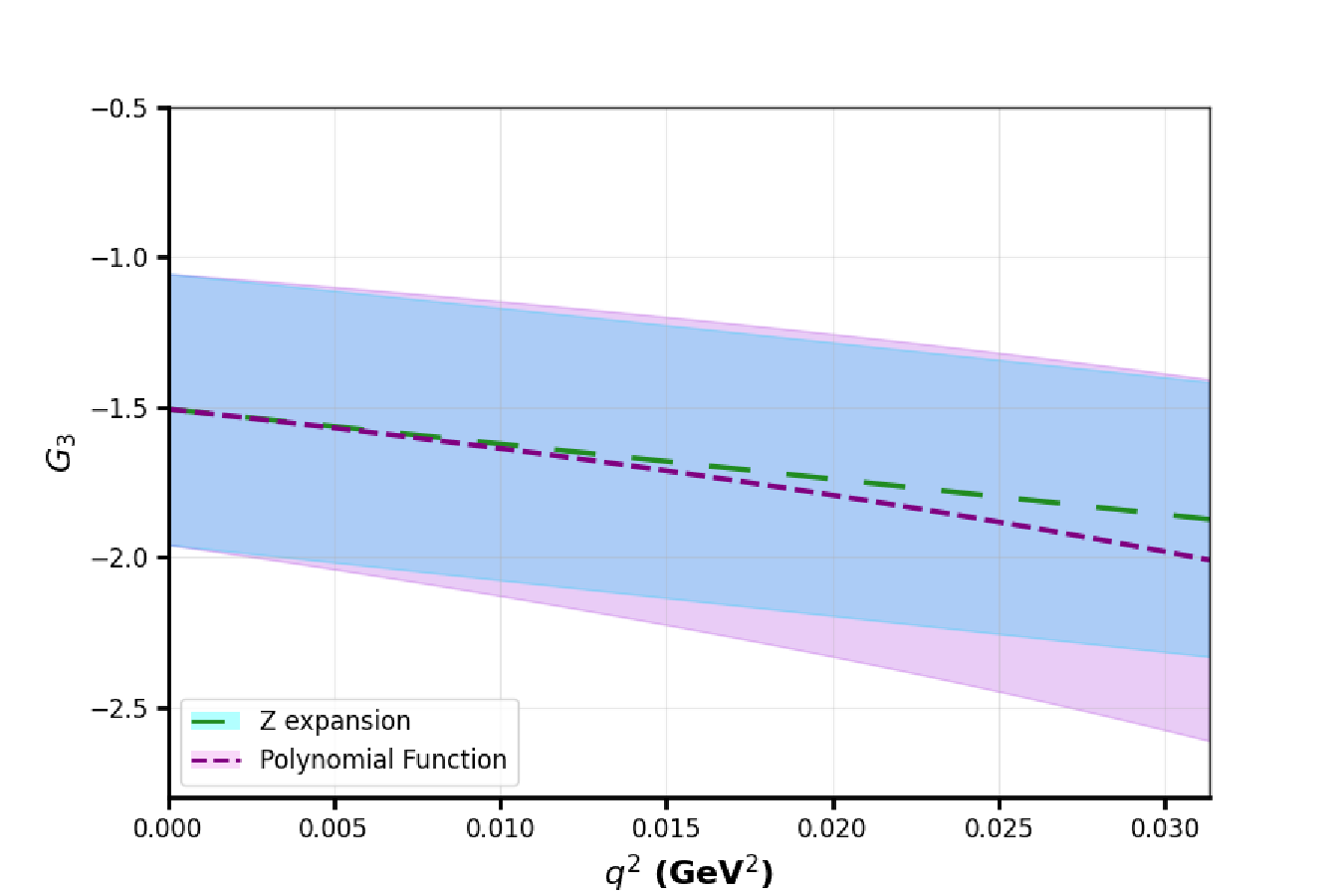}
	\caption{The $q^2$ dependence of form factors plotted with polynomial and $z$ expansion fits including error bars, using central values of auxiliary parameters.}\label{Fig:formfactorserror2}
\end{figure}

\begin{table}[h!]
	\begin{ruledtabular}
		\begin{tabular}{|c|c|c|c|c|c|c|}
					& $F_1(q^2): \slashed {p}'\gamma_{\mu}\slashed {p}$ & $F_2(q^2):p_\mu \slashed {p}'$  & $F_3(q^2):p'_{\mu}\slashed {p}$   & $G_1(q^2):  \slashed {p'} \gamma_\mu \gamma_5$ & $G_2(q^2):p_\mu \slashed {p}' \gamma_{5}$  & $G_3(q^2):p'_\mu \slashed {p} \gamma_{5}$       \\
			\hline
			$F_0(q^2=0)$ & $-0.24\pm0.08$        & $-0.11\pm0.03$      & $1.58\pm0.47$     & $-0.76{}^{+0.21}_{-022}$  & $-0.44\pm0.13$  & $-1.50\pm0.45$ \\
			$A_1$           & $43.21$          & $32.20$            &$-249.78$            & $5.40$           & $112.36$            &$ 178.87$            \\
					\end{tabular}
		\caption{Parameters for the $z$-expansion fitting of all form factors related to   $\Lambda \to p  l\bar\nu_{\ell}$.}
		\label{Tab:parameterfit2}
	\end{ruledtabular}
\end{table}
 It should be noted that the uncertainties in the values listed in Tables \ref{Tab:parameterfit1} and  \ref{Tab:parameterfit2} originate from variations in both the input and auxiliary parameters. It must also be emphasized that, as is evident from the results obtained on both the QCD and phenomenological sides, the structures presented in Tables \ref{Tab:parameterfit1} and  \ref{Tab:parameterfit2} are not unique for determining the form factors. Within the framework of the sum-rule method, the form factors are, in practice, dependent on the choice of the employed structure. In general, structures that involve more momenta exhibit greater stability with respect to variations in the auxiliary parameters. In this study, the selected structures have been determined based on the working ranges of the auxiliary parameters, thereby making uncertainties smaller in the results and ensuring that the method's requirements, as outlined in the preceding sections, are properly satisfied. Figure \ref{Fig:formfactor1} illustrates the variations of the form factors $F_1$, $F_2$, $F_3$, $G_1$, $G_2$, and $G_3$ as functions of $q^2$ within the range $ m_l^2\leq q^2 \leq (m_{\Lambda}-m_{p})^2$ for different Lorentz structures $\slashed {p}'\gamma_{\mu}\slashed {p}$, $p_\mu \slashed {p}'$, $p'_{\mu}\slashed {p}$,  $ \slashed {p'} \gamma_\mu \gamma_5$,  $p_\mu \slashed {p}' \gamma_{5}$, and  $p'_\mu \slashed {p} \gamma_{5}$, respectively. In addition to the the sum-rule results the figure includes the results obtained from polynomial fit function and the $z$ expansion, allowing for a comparison and assessment of the accuracy of different methods. Moreover, Fig \ref{Fig:formfactorserror2} presents the same variations along with the uncertainties associated with the form-factor calculations. These fitted form factors are then employed to compute the decay widths for the different lepton channels.
 \section{ Decay Widths and Branching Ratios}~\label{sec:four} 
 At this stage, using the form factors obtained in the previous section, the semileptonic decay width of the process $\Lambda \to p  l\bar\nu_{\ell}$ is calculated for the electron and muon leptonic channel. First, the general formula of the differential decay width is introduced as follows \cite{Faustov:2016pal,Gutsche:2014zna,Migura:2006en,Korner:1994nh,Bialas:1992ny}: 
 \begin{equation}\label{eq:dgamma}
	\frac{d\Gamma(\Lambda\rightarrow p  ~{\ell}\bar\nu_{\ell})}{dq^2}=\frac{G_F^2}{(2\pi)^3}
	|V_{us}|^2\frac{\lambda^{1/2}(q^2-m_\ell^2)^2}{48m_{\Lambda}^3q^2}{\cal
		H}_{tot}(q^2).
\end{equation}
 Here, $\lambda$ denotes the Källén function, which in this decay is given as follows:
 \begin{equation}\label{eq:Kallen}
 \lambda\equiv\lambda(m^2_{\Lambda}, m^2_{p}, q^2)=m^4_{\Lambda}+m^4_{p}+q^4-2(m^2_{\Lambda}m^2_{p}+m^2_{\Lambda}q^2+m^2_{p}q^2),
\end{equation}
and $m_l$ is the mass of the  lepton (electron or muon). In addition, the quantity ${\cal H}_{\rm tot}(q^2)$ representing the total helicity can be expressed as
 \begin{equation}
 	\label{eq:hh}
 	{\cal H}_{tot}(q^2)=[{\cal H}_U(q^2)+{\cal H}_L(q^2)] \left(1+\frac{m_\ell^2}{2q^2}\right)+\frac{3m_\ell^2}{2q^2}{\cal H}_S(q^2),
 \end{equation} 
 where ${\cal H}_U(q^2)$ and ${\cal H}_L(q^2)$ denote the transverse and longitudinal helicity contributions, respectively, and ${\cal H}_S(q^2)$ represents the scalar helicity component. These components can be expressed in terms of the helicity amplitudes as
\begin{eqnarray}
	\label{eq:hhc}
	&&{\cal H}_U(q^2)=|H_{+1/2,+1}|^2+|H_{-1/2,-1}|^2,\notag\\
	&&{\cal H}_L(q^2)=|H_{+1/2,0}|^2+|H_{-1/2,0}|^2,\notag\\
	&&{\cal H}_S(q^2)=|H_{+1/2,t}|^2+|H_{-1/2,t}|^2.
\end{eqnarray} 
 The helicity amplitudes are linearly related to the transition form factors $F_i$ and $G_i$: 
\begin{eqnarray}
	\label{eq:ha}
	H^{V,A}_{+1/2,\, 0}&=&\frac1{\sqrt{q^2}}{\sqrt{2m_{\Lambda}m_p(\sigma\mp 1)}}
	[(m_{\Lambda} \pm m_p){\cal F}^{V,A}_1(\sigma) \pm m_p
	(\sigma\pm 1){\cal F}^{V,A}_2(\sigma)\cr
	&& \pm m_{\Lambda} (\sigma\pm 1){\cal F}^{V,A}_3(\sigma)],\cr
	H^{V,A}_{+1/2,\, 1}&=&-2\sqrt{m_{\Lambda}m_p(\sigma\mp 1)}
	{\cal F}^{V,A}_1(\sigma),\cr
	H^{V,A}_{+1/2,\, t}&=&\frac1{\sqrt{q^2}}{\sqrt{2m_{\Lambda}m_p(\sigma\pm 1)}}
	[(m_{\Lambda} \mp m_p){\cal F}^{V,A}_1(\sigma) \pm(m_{\Lambda}- m_p \sigma
	){\cal F}^{V,A}_2(\sigma)\cr
	&& \pm (m_{\Lambda} \sigma- m_p){\cal F}^{V,A}_3(\sigma)],
\end{eqnarray}
where
\begin{equation}
\sigma=\frac{m_{\Lambda}^2+m_p^2-q^2}
{2m_{\Lambda}m_p}.
\end{equation} 
In this framework, the notation is chosen in a way that  ${\cal F}^V_i\equiv F_i$ and  ${\cal F}^A_i\equiv G_i$ (for $i=1,2,3$)  describing the vector and axial components, respectively. The plus sign corresponds to the vector current, while the minus sign corresponds to the axial current. Furthermore, $H^{V,A}_{h',\,h_W}$ shows the helicity amplitudes of the weak currents, where $h'$ is the helicity of the final baryon, and $h_W$ is the helicity of the virtual $W$ boson. The amplitudes corresponding to negative helicity values are also obtained through the following standard relations:
\begin{equation}
	H^{V,A}_{-h',\,-h_W}=\pm H^{V,A}_{h',\,h_W}.
\end{equation}
The total helicity amplitude for the  V-A current is given by
\begin{equation}
H_{h',\,h_W}=H^{V}_{h',\,h_W}-H^{A}_{h',\,h_W}.
\end{equation}
The calculated decay widths and branching ratios, along with their errors, are presented in Table~\ref{DECAY}. The ratio of decay widths in the $\mu$  to $e$  channels is also presented in  Table~\ref{DECAY}, as it reduces uncertainties. This quantity is defined as follows:
\begin{eqnarray}
	R^{\mu e}=\frac{\Gamma~[\Lambda\to p\mu~ {\overline{\nu}}_{\mu}]}{\Gamma~[\Lambda\to p e~ {\overline{\nu}}_{e}]}.
	\label{ratio}\end{eqnarray}

\begin{table}[h!]
\begin{ruledtabular}
	\begin{tabular}{|c|c|c|c|c|c|c|c|c|c|}
		  \multicolumn{5}{|c|}{polynomial  function} & \multicolumn{5}{c|}{$z$- expansion}  \\ \hline
		  $\Gamma_{e }\times10^{18}$ & $\Gamma_{\mu}  \times10^{19}$ & ${\cal B}_{e}\times10^{4}$& ${\cal B_{\mu}}\times10^{4}$ &$R^{\mu e}$& $\Gamma_{e} \times10^{18}$ & $\Gamma_{\mu} \times10^{19}$ & ${\cal B}_{e}\times10^{4}$& ${\cal B_{\mu}}\times10^{4}$&$R^{\mu e}$ \\ \hline\hline
		$1.92{}^{+0.89}_{-0.31}$ &$ 3.79{}^{+2.02}_{-0.81}$& $7.64{}^{+3.56}_{-1.25}$& $1.50{}^{+0.80}_{-0.32}$&$0.196{}^{+0.009}_{-0.012}$&$1.82{}^{+0.72}_{-0.23}$&$ 3.19{}^{+1.13}_{-0.37}$&$7.23{}^{+2.87}_{-0.94}$&$1.26{}^{+0.45}_{-0.14}$&$0.174{}^{+0.002}_{-0.005}$
		\\ 
		\end{tabular} 
	\caption{Decay widths (in GeV), branching ratios, and the parameter $R^{\mu e}$ for the muon and electron channels,  using polynomial and $z$-expansion fits.}
	\label{DECAY}
	\end{ruledtabular}
	\end{table}

\begin{table}[!h]
	\begin{tabular}{|c|c|c|c|c|c|}
			\hline\hline
					\multirow{2}{*}{} &  \multicolumn{2}{|c|}{This work} & \multicolumn{1}{c|}{PDG \cite{ParticleDataGroup:2024cfk}}& \multicolumn{1}{c|}{QCD sum rules \cite{Zhang:2024ick}}& \multicolumn{1}{c|}{lattice QCD \cite{Bacchio:2025auj}}   \\ \cline{2-6} 
		 &	\multicolumn{1}{|c|}{polynomial  function} &\multicolumn{1}{c|}{$z$- expansion} & \multicolumn{1}{c|}{}& \multicolumn{1}{c|}{$z$- expansion}& \multicolumn{1}{c|}{$z$- expansion} \\ \cline{2-6} 
	 \hline\hline
	  ${\cal B}_{e}\times10^{4}$&$7.64{}^{+3.56}_{-1.25}$&$7.23{}^{+2.87}_{-0.94}$&$8.34{}^{+0.14}_{-0.14}$&$7.72{}^{+0.64}_{-0.64}$&$7.93{}^{+0.39}_{-0.39}$\\  \hline\hline
	  ${\cal B_{\mu}}\times10^{4}$&$1.50{}^{+0.80}_{-0.32}$&$1.26{}^{+0.45}_{-0.14}$&$1.51{}^{+0.19}_{-0.19}$&$1.35{}^{+0.11}_{-0.11}$&$1.49{}^{+0.14}_{-0.14}$\\  \hline\hline
		   	 	  	\end{tabular}
	\caption{Comparison of branching ratios for the electron and muon channels, with results obtained from lattice calculations \cite{Bacchio:2025auj}, QCD sum rules \cite{Zhang:2024ick}, and PDG data \cite{ParticleDataGroup:2024cfk}.}
	\label{results}
\end{table}

As shown in Table~\Ref{results}, we compare our branching ratio results with those reported by the PDG \cite{ParticleDataGroup:2024cfk}, as well as theoretical calculations in the literature \cite{Zhang:2024ick, Bacchio:2025auj}. The values of branching ratios obtained in the present work using the polynomial expansion are consistent within uncertainties with the reference values reported by the PDG \cite{ParticleDataGroup:2024cfk} (see Table~\Ref{results}), with the muon channel showing a central value close to that of PDG. Similarly, the results from the $z$-expansion approach are consistent with the PDG values within uncertainties for both the muon and electron channels. This observation indicates that the polynomial form appears to provide a reasonable description of the relevant form factors in the extraction of branching ratios, while emphasizing that the central values should be interpreted with caution when the uncertainties are relatively large.
   The branching ratios obtained in the present work are also consistent within      uncertainties with those from lattice QCD calculations \cite{Bacchio:2025auj} and with results from QCD sum rules \cite{Zhang:2024ick}, as summarized in Table~\Ref{results}. This agreement further supports the overall consistency of our results. These findings suggest that the present approach is reliable for phenomenological analyses of leptonic decay channels, without overemphasizing the significance of the central values.
According to Eq.~(\ref{ratio}), the  ratio of decay widths in the $\mu$  to $e$  channels calculated in the present work using the polynomial function is $R^{\mu e} = 0.196^{+0.009}_{-0.012}$, while using the $z$ expansion it is $R^{\mu e} = 0.174^{+0.002}_{-0.005}$. This  quantity obtained from the $z$ expansion, in particular, shows very good agreement with the value reported by the BESIII Collaboration \cite{BESIII:2021ynj}, $R^{\mu e} = 0.178^{+0.028}_{-0.028}$.  In lattice QCD, it is found to be $R^{\mu e} = 0.188^{+0.008}_{-0.008}$ ~\cite{Bacchio:2025auj}. We note that the uncertainties in our results are slightly larger than those in Ref. \cite{Zhang:2024ick}, 
mainly because we adopt wider variation ranges for the Borel parameters 
to ensure the dominance of the pole contribution and the convergence of the OPE. 
This more conservative treatment leads to larger but more reliable error estimates. These comparisons indicate that our results are in very good agreement with both theoretical and experimental determinations.

 \section { SUMMARY AND CONCLUSION }\label{sec:five}

The investigation of the internal structure, intrinsic properties, and decay mechanisms of light baryons has attracted increasing interest in both theoretical and experimental research, particularly as experimental evidence continues to validate their existence. Among the key areas in light-baryon physics is the study of their decay processes, especially weak transitions, which provide important opportunities to probe potential signals of physics beyond the SM. 
In this work, we employed the standard QCD sum-rule approach with the relevant three-point correlation function to investigate the hyperon semileptonic decay $\Lambda \to p\,\ell\,\bar{\nu}_{\ell}$, incorporating nonperturbative operators up to dimension five and appropriately chosen Borel and other auxiliary parameters. We determined all six form factors, three vector and three axial vector, for the semileptonic decay $\Lambda \to p\,\ell\,\bar{\nu}_{\ell}$ at $q^{2}=0$. The complete set of form factors was obtained using both polynomial fits and the $z$ expansion method, from which we extracted their $q^{2}$ dependence for different Lorentz structures. These fitted form factors were further applied to calculate the decay widths and branching fractions of the electron and muon channels.  We compared the results for the  branching ratios for both lepton channels with the existing results from the experiment and different theoretical approaches including lattice QCD.  Overall, we observed a good consistency of our predictions and the experimental and theoretical results of other approaches on the branching fractions. For the semileptonic decay under consideration, we obtained $R^{\mu e} = 0.196^{+0.009}_{-0.012}$ using the polynomial fit and $R^{\mu e} = 0.174^{+0.002}_{-0.005}$ using the $z$ expansion. The latter provides significantly higher precision, and our findings show excellent consistency with the values reported by the PDG. Being experimentally accessible, this semileptonic decay, involving the underlying quark-level transition $s \to u$ serves as a benchmark for studying hyperon semileptonic transitions and for exploring potential signals of physics beyond the SM.

\section*{APPENDIX: DETAILS OF EXPRESSIONS ON THE QCD SIDE OF THE SUM RULES}

The detailed formulas for both the perturbative and nonperturbative parts of the correlation function on the QCD side, corresponding to the semileptonic decay process $\Lambda \to p \ell \bar{\nu}_\ell$ and associated with the  $ \slashed {p'} \gamma_\mu \gamma_5$ structure, are presented as follows:

\begin{eqnarray} 
\rho^{\mathrm{Pert.}}_{p_\mu \slashed {p}' \gamma_{5}}(s,s',q^2)&=&\frac {1}{64 \sqrt{6} \pi^4 Z_1}\Bigg\{ \frac12(3 m_s (5 m_d - 3 m_u) m_u v + \big[-42 m_d m_s m_u - m_u v (2 s u + S_1 v) \notag\\ &+ &
    m_s (7 s u^2 + 6 S_1 u v + 5 s' v^2)\big] Z_1
   + 15 (m_d - m_u) s u Z_1^2 + 
 \Delta \big[7 m_u v + 3 m_s (4 u + 5 v) + 18 m_d Z_1 - 
    12 m_u Z_1)\big] \notag\\
   &+&\beta \bigg(3 m_s m_u (-4 m_d + m_u) v + 
   m_s \big(12 m_d m_u + 3 s u^2 + 2 S_1 u v + s' v^2\big) Z_1 + 
   3 (m_d - 2 m_u) s u Z_1^2\notag\\
   & + &
   \Delta \big(4 m_s u + 3 m_s v + m_u v + 3 m_d Z_1 - 15 m_u Z_1\big)\bigg)+\beta^2 \bigg(15 m_s m_u (m_d + m_u) v \notag\\ &+& \big(78 m_d m_s m_u + 
      9 m_s s u^2 + (8 m_u s + 11 m_s S_1) u v + (4 m_u S_1 + 
         13 m_s s') v^2\big) Z_1 + 3 (5 m_d + m_u) s u Z_1^2 \notag\\&+ &
   \Delta (20 m_s u + 39 m_s v + 21 m_u v + 12 (2 m_d + m_u) Z_1)\bigg) \Bigg\}\Theta[\Delta(s,s^{\prime},q^2)],
    	\end{eqnarray}

\begin{eqnarray} 
\rho^{\mathrm{Dim3}}_ {p_\mu \slashed {p}' \gamma_{5}}(s,s',q^2)&=&\frac {1}{8 \sqrt{6} \pi^4 }\Bigg\{ -(3 \langle \bar{d} d\rangle- 2 \langle \bar{u} u\rangle) Z_1-\beta ( \langle \bar{d} d\rangle - 5  \langle \bar{u} u\rangle) Z_1-\beta^2 (2 \langle \bar{d} d\rangle +  \langle \bar{u} u\rangle) Z_1\Bigg\} \Theta[\Delta(s,s^{\prime},q^2)],
    	\end{eqnarray} 

\begin{eqnarray} 
\rho^{\mathrm{Dim4}}_ {p_\mu \slashed {p}' \gamma_{5}}(s,s',q^2)&=&0,
    	\end{eqnarray} 
and
\begin{eqnarray} 
\tilde{\Gamma}^{\mathrm{Dim5}}_ {p_\mu \slashed {p}' \gamma_{5}}(s,s',q^2)&=&0,
    	\end{eqnarray} 

	where
	\begin{eqnarray}
	\Delta(s,s',q^2) &=&q^2 u v + s u Z_1 + s' v Z_1.
	\end{eqnarray}	
	$\Theta[\Delta(s,s^{\prime},q^2)]$ denotes the unit step function, with the subsequent relations applied:	
	
	\begin{eqnarray}
	Z_1 &=&(1 - u - v),\nonumber\\
	S_1&=&(-q^2 + s + s').
\end{eqnarray}	
	

\end{document}